\documentclass[aps,prc,preprint,superscriptaddress,showpacs]{revtex4}
\usepackage[dvipdfmx]{graphicx}
\usepackage{amsmath}
\renewcommand{\theequation}
{{\rm \thesection.\arabic{equation}}}
\newcommand{\bm}[1]{\mbox{\boldmath $ #1 $}}
\def\be{\begin{equation}}
\def\ee{\end{equation}}
\def\bea{\begin{eqnarray}}
\def\eea{\end{eqnarray}}

\def\eps{\varepsilon}
\def\eps0{\varepsilon_0}

\begin{document}
\title{Formulation of effective interaction in terms of renormalized 
vertices and propagators}
\author{Kenji Suzuki}
\email[]{suz93@mocha.ocn.ne.jp}
\affiliation{Senior Academy, Kyushu Institute of Technology, 
             Kitakyushu 804-8550, Japan}
\author{Hiroo Kumagai}
\email[]{kumagai@fit.ac.jp}
\affiliation{Faculty of Information Engineering, Fukuoka Institute of Technology, 
             Fukuoka 811-0295, Japan}
\author{Masayuki Matsuzaki}
\email[]{matsuza@fukuoka-edu.ac.jp}
\affiliation{Department of Physics, Fukuoka University of Education, 
             Munakata, Fukuoka 811-4192, Japan}     
\author{Ryoji Okamoto}
\email[]{okamoto.ryoji.munakata@gmail.com}
\affiliation{Senior Academy, Kyushu Institute of Technology, 
             Kitakyushu 804-8550, Japan}
                                                
\date{\today}
\begin{abstract}
One of the useful and practical methods for solving quantum-mechanical 
many-body systems is to recast the full problem into a form of the effective 
interaction acting within a model space of tractable size.
Many of the effective-interaction theories in nuclear physics have been formulated by use of 
the so called $\widehat Q$ box introduced by Kuo et al.
It has been one of the central problems how to calculate the $\widehat Q$ box 
accurately and efficiently. 
We first show that, introducing new basis states, the Hamiltonian  is transformed 
to a block-tridiagonal form in terms of submatrices with small dimension.
With this transformed Hamiltonian, we next prove that the $\widehat Q$ box 
can be expressed in two ways: 
One is a form of continued fraction and the other is a simple series expansion 
up to second order with respect to renormalized vertices and propagators. 
This procedure ensures to derive an exact $\widehat Q$ box, if the
calculation converges as the dimension of the Hilbert space
tends to infinity. The $\widehat Q$ box given in this study corresponds to
a non-perturbative solution for the energy-dependent effective
interaction  which is often referred to as the Bloch-Horowitz or
the Feshbach form.
By applying the $\widehat Z$-box approach based on the $\widehat Q$ box proposed previously,
we introduce a graphical method  for solving 
the eigenvalue problem of the Hamiltonian.  The present approach has a possibility
of resolving many of the difficulties encountered in the effective-interaction theory.
\end{abstract}
\maketitle
%
\section{Introduction}
In nuclear many-body physics various methods have been proposed, on the basis of the shell model, 
 to solve the Schr\"{o}dinger 
equations for  nuclear many-body systems starting with realistic nucleon-nucleon interactions.   
These methods, which are called the {\it ab initio} calculations, include the Green's function  
Monte Carlo (GFMC) method \cite{PRWC01, PW01}, the no-core shell model (NCSM) 
\cite{NVB00-2, LBKNSV08}, the effective interaction  for hyperspherical harmonics (EIHH) method
\cite{LO12},
the coupled cluster method (CCM) \cite{BM07, HPDH08, HPDH09}
and the unitary-model-operator approach (UMOA) \cite{SOK87, FOS04, FOS09}.
Much effort has been made also to diagonalize a matrix of a many-body shell-model 
Hamiltonian in a huge dimensional Hilbert space on the basis of , or alternatively 
to the Lanczos method \cite{Lan50, GL96, MKHS10}.

The shell model calculations were carried out in the early stage 
by introducing the phenomenological residual interaction between two nucleons determined 
from the experimental data \cite{Tal63, MBZ64}. 
These studies have been considered to be useful in accounting 
the variety of nuclear properties, of which studies were reviewed by Talmi \cite{Tal03}. 
The next stage of the 
nuclear shell-model calculation was to employ a realistic nucleon-nucleon (NN) interaction and to derive 
theoretically a renormalized interaction which takes the repulsive-short-range correlations into account. 
The first attempt of this approach was made by Dowson, Talmi and Walecka \cite{DTW62} 
by applying the Brueckner 
reaction-matrix theory. Soon afterwords the correction to the reaction matrix, such as the 
core-polarization effect, was estimated by Bertsch \cite{Ber65}.

A marked development was attained by Kuo and Brown \cite{KB66}, who performed a second-order perturbative 
calculation for deriving the effective interaction between two valence nucleons outside the core 
$^{16}$O. They have established that the core-polarization effect has a crucial role in understanding 
the nuclear properties. Their study took an increasing attention to the evaluation of higher-order 
perturbative terms. The third-order diagrams were calculated by Barrett and Kirson \cite{BK70} 
and many studies were made 
to sum up the specific series of diagrams to all orders, which include the Pad\`{e} 
approximants \cite{HSK76, AM79}, RPA \cite{Kir74} 
and the induced-interaction method\cite{Kir71, BB73, HHKBB05} . The theoretical formalism for deriving 
the effective interaction 
was also developed on the basis of the perturbation theory.  The folded-diagram theory by Kuo, Lee 
and Ratcliff \cite{KLR71} was proposed and has been recognized to be the basic formalism of deriving microscopically 
the effective interaction. Much effort has been devoted continuously to the progress in the 
effective-interaction theory and its practical application\cite{Bra67,BK73,EO77,KO90,And87,RW12}. 
The present status of these studies were 
reviewed by recent articles of Coraggio et al \cite{CCGIK12, CCGIK09}.
This effective-interaction method has been developed to apply to new fields of many-body physics 
such as quantum dots \cite{VNUS01, PHHKP11} and many-boson systems \cite{CFAR09}.

Most of the effective-interaction theories given to date have been formulated in terms 
of the $\widehat Q$ box introduced by Kuo and his collaborators \cite{KO90, HKO95,DEHKO04}.  
Originally the $\widehat Q$ box has been defined as the sum of linked and unfolded diagrams 
\cite{KLR71}. 
In the algebraic or non-diagrammatical approach the $\widehat Q$ box is 
equivalent to the energy-dependent effective interaction given by 
Bloch and Horowitz \cite{BH58} and Feshbach \cite{Fesh62} which has been studied extensively on the 
Brillouin-Wigner perturbation theory \cite{LM86,WH12}.

It has been established that the effective interaction can be expressed as a series expansion 
in terms of the $\widehat Q$ box and its energy derivatives. 
The series can be summed up by using either the Krenciglowa-Kuo (KK) \cite{KK74} or 
the Lee-Suzuki(LS) \cite{LS80, SL80,SOEK94} methods.  
It has been known that, in general, two methods have different convergence properties: 
Many of the numerical calculations have shown that the KK method yields the eigenvalues 
for the eigenstates which have the largest overlaps with the chosen model space.  
However, it has been pointed out that the rigorous convergence condition for the KK method 
has not yet been clarified \cite{Tak11b}.
On the other hand the LS method reproduces the eigenvalues which lie closest 
to the chosen unperturbed energy.
Both of the two approaches reproduce only certain of the eigenvalues of the original Hamiltonian.  
This restriction is not, in general, desirable.

Another difficulty encountered in actual calculations is the pole problem. 
The $\widehat Q$ box itself has poles at the energies which are the eigenvalues of 
$QHQ$, where $Q$ is the projection operator onto the complement ($Q$ space) 
of the model space ($P$ space).
The presence of the poles causes often instability in numerical calculations.
Three of the present authors and Fujii \cite{SOKF11} have shown that it was indeed possible to resolve these
difficulties by introducing a new vertex function $\widehat Z(E)$, called the $\widehat Z$ box.
The $\widehat Z$-box approach based on the $\widehat Q$ box may have 
a possibility of resolving many of the difficulties encountered in the effective-interaction theory. 

At present the most important remaining task would be to establish a method of how to calculate 
the $\widehat Q$ box rigorously and efficiently.  
The perturbative calculation method for the $\widehat Q$ box has been established 
and applied widely\cite{KLR71,KK74,KO90}. 
In the derivation of the nuclear effective 
interaction, the convergence of the order-by-order calculation was confirmed in many of 
the numerical studies \cite{CCGIK09,CCGIK12}. 
However, a basic problem of the convergence of its perturbation expansion 
has not been made clear theoretically for general cases.
Main concern of the present study is to propose a non-perturbative 
method for obtaining a convergent result for any of the starting NN interactions.

The formulation in the present study consists mainly of two parts: 
The first one is to transform the Hamiltonian to a block-tridiagonal form, where the dimensions 
of the block submatrices are taken to be equal-to-or-less-than the dimension of the $P$ space.
With the block-tridiagonalized Hamiltonian, the next step is to derive a set of coupled equations 
for determining the $\widehat Q$ box.
We show that the coupled equations can be solved by employing two different 
recursion methods: 
The first solution is represented in a form of continued fraction, and the second one is expressed 
as a sum of terms up to second order with respect to renormalized vertices and propagators. 
In both of the methods the calculation of the $\widehat Q$ box can be carried out 
without matrix inversion of $QHQ$ which is usually a huge-dimensional matrix. 
All the procedures for obtaining the $\widehat Q$ box are reduced to calculations of small-dimensional 
submatrices in the block-tridiagonalized Hamiltonian.
 
Regarding the block tridiagonalization of the Hamiltonian, the present approach has 
a common aspect to the so called block Lanczos method based on the theory 
of the Krylov subspaces \cite{GL96}.
For a given model space $P$ and a Hamiltonian $H$, the subspaces leading 
to a block-tridiagonal form of $H$ are determined uniquely. 
Therefore, the subspaces given in the present study are the same as those of Krylov.
However, the choice of basis states of each subspace is ambiguous.
For determining the basis states we employ a different calculation procedure from the usual one 
in the block Lanczos method.
Different basis states are introduced, and we show that they are suitable for the purpose of 
calculating not only the $\widehat Q$ box but also the eigenstates of $H$. 

The construction of the present article is as follows: In Section II some basic elements 
of the effective-interaction theory are reviewed. 
Section III is devoted to the formulation of rigorous calculation of the $\widehat Q$ box.
A set of coupled equations for determining the $\widehat Q$ box are given.
The equations are solved by employing recursion methods and two kinds of
solutions for the $\widehat Q$ box are derived.
In Section IV, a method is given for the problem of how to calculate eigenstates of $H$ 
within the framework of the effective-interaction theory.
In Section V a short review of the $\widehat Z$-box theory is given.
In Section VI, by applying the $\widehat Z$-box theory, we make a numerical calculation 
with a model Hamiltonian to assess the present approach.
We propose a graphical method and show that it works well for finding the eigenvalues of $H$.
Summary of the present study and some remarks are given in the last Section.
In Appendices A and B the derivatives of the $\widehat Q$ box are given 
for the two recursive solutions, which are necessary for calculating the $\widehat Z$ box.

%
%
\section{Effective-interaction theory by means of similarity transformation}
\setcounter{equation}{0}
Let us begin with a Hamiltonian $H$ defined in a Hilbert space. We divide 
the space into a model space ($P$ space) and its complementary space 
($Q$ space). When all the eigenvalues of an operator $H_{\rm eff}$ given in the 
$P$ space coincide with those of $H$, we call $H_{\rm eff}$ an effective 
Hamiltonian. 
In the following, we do not impose any particular conditions on $H$ and states 
belonging to the $P$ space nor assume degeneracy of their unperturbed energies.  

There are various ways of constructing $H_{\rm eff}$. 
We adopt the following standard one. 
First we introduce an operator $\omega$ that maps states in the 
$P$ space and those in the $Q$ space to each other, with the properties 
\cite{SL80}, 
\begin{align}
\label{eq:omega-definition}
 \omega &= Q\omega P, \\
 \label{eq:omega-property}
 \omega^{n} &= 0 \hspace{10mm} (n\ge 2).
\end{align}
The operator $\omega$ defines a similarity transformation of $H$, 
\begin{align}
\label{eq:H-transformation}
 \widetilde{H} &= {\rm e}^{-\omega}H {\rm e}^{\omega}.
\end{align}
This reduces to 
\begin{align}
 \label{eq:transformed -hamiltonian}
 \widetilde{H}&= (1-\omega)H(1+\omega)
\end{align}
by virtue of Eq.(\ref{eq:omega-property}). 

The condition that $P\widetilde{H}P$ be a model-space effective Hamiltonian $H_{\rm eff}$ is that $\widetilde{H}$ 
should be decoupled between the $P$ and $Q$ spaces as 
\begin{align}
\label{eq:decoupling-eq1}
 Q\widetilde{H} P &=  0.
\end{align}
This condition is rewritten as
\begin{align}
\label{eq:decoupling-eq2}
 QHP+QHQ\omega-\omega PHP-\omega PHQ\omega&= 0,
\end{align}
with the aid of Eqs.(\ref{eq:omega-definition}) and (\ref{eq:transformed -hamiltonian}). 
This equation for $\omega$ was first derived by Okubo  \cite{Oku54} in a different way. 
Once a solution $\omega$ to Eq.(\ref{eq:decoupling-eq2}) is given, $H_{\rm eff}$ is written as 
\begin{align}
\label{eq:eff-Hamiltonian}
 H_{\rm eff}&= PHP+PHQ\omega.
\end{align}
Dividing $PHP$ into the unperturbed part $PH_0P$ and the interaction $PVP$, we write 
\begin{align}
\label{eq:divide-PHP}
    PHP= PH_{0}P + PVP.
\end{align}
The model-space effective interaction $V_{\rm eff}$ is defined as
\begin{align}
\label{eq:definition-Veff}
    V_{\rm eff} &= H_{\rm eff} - PH_0P \nonumber\\
         &= PVP + PHQ\omega .
\end{align}
From the definition of $H_{\rm eff}$ and $V_{\rm eff}$ we see that a central part 
of determining them is to find a solution for $\omega$ in Eq.(\ref{eq:decoupling-eq2}).

Since Eq.(\ref{eq:decoupling-eq2}) is a nonlinear matrix equation for $\omega$, it is 
difficult to find a general solution. 
The following formal solution, however, has been known and is enough for applications. 
We rewrite 
Eq.(\ref{eq:decoupling-eq2}) as 
\begin{align}
\label{eq:decoupling-eq3}
 QHP+ QHQ\omega - \omega H_{\rm eff}&= 0,
\end{align}
using Eq.(\ref{eq:eff-Hamiltonian}).  
Here the eigenvalue equation for $H_{\rm eff}$ is given by
\begin{align}
\label{eq:eff-Hamiltonian-eigenvalue-eq}
 H_{\rm eff}|\phi_{k}\rangle &= E_{k}|\phi_{k}\rangle.
\end{align}
If the operator $\omega$ is a solution to Eq.(\ref{eq:decoupling-eq2}), we can verify that
the eigenstates $\{|\phi_{k}\rangle\}$ belong to the $P$ space and each eigenvalue 
$E_{k}$ coincides with one of those of $H$. 
The effective Hamiltonian $H_{\rm eff}$ is not Hermitian in general; 
the eigenstates $\{ |\phi_{k}\rangle, k=1, 2,\cdots, d\}$ are not 
orthogonal to each other. Then we introduce the adjoint states 
$\{ \langle \widetilde{\phi}_{k}|, k=1, 2,\cdots, d\}$ according to the biorthogonality condition 
\begin{align}
\label{eq:bi-orthogonal-state}
 \langle \widetilde{\phi}_{k}|\phi_{k'}\rangle &= \delta_{kk'},
\end{align}
where $d$ is the dimension of the $P$ space. The projection operator onto the 
$P$ space is written as 
\begin{align}
\label{eq:P-space-projection-operator}
 P &= \displaystyle{\sum_{k=1}^{d} |{\phi}_{k}\rangle \langle\widetilde{\phi}_{k}|}.
\end{align}

Then, using Eqs.(\ref{eq:decoupling-eq3})--(\ref{eq:P-space-projection-operator}), $\omega$ is given by 
\begin{align}
\label{eq:omega-representation}
 \omega &=\displaystyle{\sum_{k=1}^{d} 
                      \frac{1}{E_{k}-QHQ}QHP
                      |\phi_{k}\rangle \langle\widetilde{\phi}_{k}|},
\end{align}
and from Eq.(\ref{eq:eff-Hamiltonian}) $H_{\rm eff}$ becomes
\begin{align}
\label{eq:tilde-H-representation-1}
 H_{\rm eff}&= PHP+\displaystyle{\sum_{k=1}^{d} 
                           PHQ\frac{1}{E_{k}-QHQ}QHP
                            |\phi_{k}\rangle \langle\widetilde{\phi}_{k}|}.
\end{align}
Here we introduce an operator in the $P$ space called the $\widehat Q$ box 
\begin{align}
\label{eq:Q-box}
 \widehat{Q} (E)&= PHP+
                           PHQ
                           \frac{1}{E-QHQ}QHP,
\end{align}
where $E$ is an energy variable. 
The $\widehat{Q}$ box thus defined is equivalent to the energy-dependent
effective Hamiltonian referred to as the Bloch-Horowitz\cite{BH58}
and /or the Feshbach\cite{Fesh62} forms.
In terms of $\widehat{Q}(E)$, $H_{\rm eff}$ is 
expressed as 
\begin{align}
\label{eq:tilde-H-representation-2}
 H_{\rm eff} &= \displaystyle{\sum_{k=1}^{d}\widehat{Q} (E_{k})
                            |\phi_{k}\rangle \langle\widetilde{\phi}_{k}|},
\end{align}
from which the following self-consistent equation can be derived
\begin{align}
\label{eq:0802a}
 \widehat Q(E_k)|\phi_{k}\rangle = E_k|\phi_{k}\rangle .
\end{align}
The $H_{\rm eff}$ in Eq.(\ref{eq:tilde-H-representation-2}) is just a formal solution in the sense 
that unknown $E_k$, 
$|\phi_k\rangle$, and $\langle\widetilde{\phi}_k|$ appear on the right-hand side, 
but the following method of solving is available: 
In order that the solutions to Eq.(\ref{eq:eff-Hamiltonian-eigenvalue-eq}) coincide 
with those given by Eq.(\ref{eq:tilde-H-representation-2}), they selfconsistently 
satisfy the iterative equation 
\begin{align}
\label{eq:Q-box-eigenvalue-eq}
 \widehat{Q}(E_{k}^{(n)})|\phi_{k}^{(n+1)}\rangle
          &= E_{k}^{(n+1)}|\phi_{k}^{(n+1)}\rangle,
\end{align}
where $E_{k}^{(n+1)}$ and $|\phi_{k}^{(n+1)}\rangle$ are the $(n+1)$-th 
order eigenvalue and eigenstate of the $\widehat Q$ box, respectively, given by the $n$-th 
order eigenvalue $E_{k}^{(n)}$. 
There have been a lot of studies about the convergence of this iterative method 
\cite{KK74,LS80,SL80,SOEK94,Tak11b}.
But the condition of convergence is rather complicated and it has been known that only 
some specific solutions are obtained. 

In addition, $\widehat{Q}(E)$ has poles at energies $\{\varepsilon_q\}$, where $\varepsilon_{q}$ is one of the eigenvalues 
of $QHQ$, 
\begin{align}
\label{eq:QHQ-eigenvalue-eq}
 QHQ|q\rangle&=\varepsilon_q |q\rangle.
\end{align}
These singularities of the $\widehat Q$ box lead to some difficulties in numerical calculations
\cite{SOKF11}. 
These arguments suggest that some further improvements are desired 
for the $\widehat Q$-box method although it has been applied widely to practical problems. 
%
%
\section{Calculation of the $\widehat Q$ box by means of recurrence relations}
\setcounter{equation}{0}

Most of the effective-interaction theories formulated so far are based on the $\widehat Q$ box. 
The $\widehat Q$ box has been calculated on the perturbative expansion methods, 
but their convergence properties and accuracies have not been well understood yet. 
This is because, as a matter of fact, it is impossible to solve  the eigenvalue problem 
of $QHQ$ nor to calculate the inverse of $(E-QHQ)$ when the dimension of the $Q$ space is huge. 
The accuracy of the $\widehat Q$ box determines that of $H_{\rm eff}$ and $V_{\rm eff}$, 
because errors that arise in the calculations of operators and/or matrices  in the $P$ space 
with small dimension are considered to be negligible. 

In the following subsections we describe a method of how to calculate accurately and efficiently 
the $\widehat Q$ box.    
We first transform $H$ to a block-tridiagonal form.
With this transformed Hamiltonian we derive a set of coupled equations for determining 
the operator $\omega$.
We shall show that these coupled equations can be solved in two ways by introducing 
two types of recurrence relations. 
The properties of two solutions for the $\widehat Q$ box are discussed.
%
\subsection{Block tridiagonalization of Hamiltonian}

\begin{figure}[h]
   \begin{center}
   \includegraphics[width=8cm]{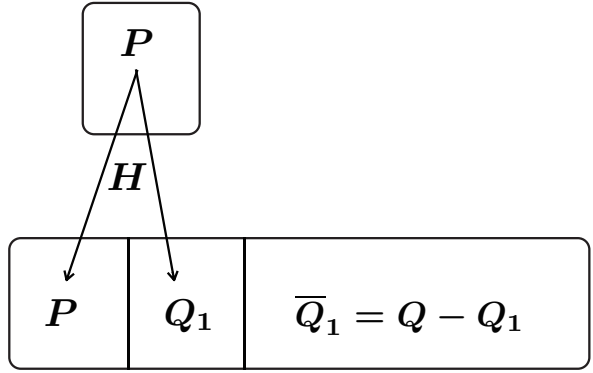}
   \caption{The image $H(P)$ by the mapping $H$.
Here, $H$ denotes the Hamiltonian, $P$ the model space and $Q_1$ the $Q$-space part of the image $H(P)$.
}
  \label{fig : mapping-P}
    \end{center}
\end{figure}

We transform the Hamiltonian $H$ into a tractable form by changing basis vectors. 
First we introduce 
\begin{align}
\label{eq:Yp-operator}
 Y_{ P} &=  PHQ\cdot QHP.
\end{align}
The $Y_{ P}$ is an operator in the $P$ space, which is Hermitian and positive 
semi-definite, that is, $y_{k}^{(1)}\ge 0$ in the eigenvalue equation
\begin{align}
\label{eq:Yp-operator-eigenvalue-eq}
 Y_{P} |p_k\rangle &= y_{k}^{(1)}|p_k\rangle. 
\end{align}
Suppose that $d_1$ eigenvalues are nonzero among $\{ y_k^{(1)}\}$. 
In terms of the eigenvectors 
$\{|p_k\rangle, k=1, 2, \cdots, d_1 \}$ with nonzero eigenvalues, we define normalized vectors 
$\{|q_k^{(1)}\rangle \}$ in the $Q$ space as
\begin{align}
\label{eq:q_k-state-definition}
 |q_{k}^{(1)}\rangle& =  \frac{1}{\sqrt{y_{k}^{(1)}}}QHP
 |p_k\rangle, \,\,\,\,(k=1, 2, \cdots, d_1).
\end{align}
They are orthogonal to each other and span the $d_1$-dimensional subspace 
$Q_1$ in the $Q$ space. 
Then the projection operator onto the $Q_1$ space becomes 
\begin{align}
\label{eq:Q1-projection-operator}
 Q_1 &= \sum_{k=1}^{d_1}|q_{k}^{(1)}\rangle\langle q_{k}^{(1)}|.
\end{align}
The complement of the $Q_1$ space in the $Q$ space is given by 
\begin{align}
\label{eq:overline-Q1}
 \overline{Q}_1&=Q-Q_1.
\end{align}
Equation (\ref{eq:q_k-state-definition}) indicates that 
\begin{align}
\label{eq:QHP-1}
 QHP &= 
      \sum_{k=1}^{d_1}\sqrt{y_k^{(1)}}|q^{(1)}_{k}\rangle \langle p_{k}|, 
\end{align}
then we have 
\begin{align}
\label{eq:QHP-2}
 QHP&=Q_{1}HP
\end{align}
which leads to 
\begin{align}
\label{eq:Q1HP-decoupling}
 \overline{Q}_{1}HP&=0.
\end{align}
Thus the image $H(P)$ by the mapping $H$ is given as a sum of the $P$ and $Q_1$ spaces 
as depicted in Fig.{\ref{fig : mapping-P}}. 
%
%

Next, a similar manipulation with replacing $P$ and $Q$ with $Q_1$ and 
$\overline{Q}_1$, respectively, leads to another orthogonal system. 
We introduce 
\begin{align}
\label{eq:YQ1}
 Y_{Q_1}&=Q_1 H \overline{Q}_1 \cdot \overline{Q}_1HQ_1,
\end{align}
and write its eigenvalue equation as 
\begin{align}
\label{eq:Yq1-operator-eigenvalue-eq}
 Y_{Q_1} |q'^{(1)}_k\rangle &= y_{k}^{(2)}|q'^{(1)}_k\rangle.
\end{align}
The eigenvectors $\{|q'^{(1)}_k\rangle \}$ belong to the $Q_1$ space 
and accordingly are given as linear combinations of 
$\{ |q^{(1)}_k \rangle,k=1, 2, \cdots,d_1\}$ in Eq.(\ref{eq:q_k-state-definition}). 
Suppose also that $d_2$ eigenvalues are nonzero among $\{ y_k^{(2)}\}$.
New orthogonal bases 
\begin{align}
\label{eq:q(2)_k-state-definition}
 |q_{k}^{(2)}\rangle & =
    \frac{1}{\sqrt{y_{k}^{(2)}}}\overline{Q}_1HQ_1
          |q'^{(1)}_k\rangle, \hspace{1cm} k=1, 2, \cdots, d_2
\end{align}
are derived. 
The $d_2$-dimensional subspace $Q_2$ is defined by them and the projection 
operator onto the $Q_2$ space is expressed as 
\begin{align}
\label{eq:Q2-projection-operator}
 Q_2 &= \sum_{k=1}^{d_2}|q_{k}^{(2)}\rangle\langle q_{k}^{(2)}|.
\end{align}
The projection operator $Q_2$ has the properties 
\begin{align}
\label{eq:Q2HP-decoupling}
 Q_2 HP  &= 0,\\
\label{eq:overQ1HQ1} 
  \overline{Q}_1 HQ_1 &= Q_2 HQ_1,\\
 \label{eq:Q2HQ1-decoupling}
   \overline{Q}_2 HQ_1 &= 0,
\end{align}
where $\overline{Q}_2$, the complementary space to $Q_1 + Q_2$ in the $Q$ space, 
is written as 
\begin{align}
\label{eq:Q_2-def}
 \overline{Q}_2 &= Q - Q_1 - Q_2.
\end{align}

Repeating these manipulations leads to the following: 
Decompose the $Q$ space as 
\begin{align}
\label{eq:Q-decomposition}
 Q &= Q_1+Q_2+\cdots +Q_n + \cdots.
\end{align}
Basis vectors of a subspace $Q_m$, namely, 
   $\{ |q^{(m)}_k \rangle,k=1, 2, \cdots,d_m\}$, 
define the projection operator 
\begin{align}
\label{eq:Qm-projection-operator}
 Q_m &=
    \sum_{k=1}^{d_m}|q_{k}^{(m)}\rangle\langle q_{k}^{(m)}|.
\end{align}
The basis vectors $\{ |q^{(m)}_k \rangle \}$ are given as follows: 
Introduce $Y_{Q_{m-1}}$ as 
\begin{align}
\label{eq:YQ(m-1)}
 Y_{Q_{m-1}} &=
        Q_{m-1} H \overline{Q}_{m-1} \cdot \overline{Q}_{m-1}HQ_{m-1}
\end{align}
with 
\begin{align}
\label{eq:Q(m-1)-decomposition}
 \overline{Q}_{m-1} &= Q-(Q_1+Q_2+\cdots +Q_{m-1}).
\end{align}
Its eigenvalue equation is 
\begin{align}
\label{eq:YQ(m-1)-operator-eigenvalue-eq}
 Y_{Q_{m-1}} |q'^{(m-1)}_k\rangle &= y_{k}^{(m)}|q'^{(m-1)}_k\rangle.
\end{align}
In general new orthogonal bases 
\begin{align}
\label{eq:q(m)_k-state-definition}
 |q_{k}^{(m)}\rangle & =  \frac{1}{\sqrt{y_{k}^{(m)}}}\overline{Q}_{m-1}HQ_{m-1}
 |q'^{(m-1)}_k\rangle
\end{align}
are derived from the eigenvectors $\{|q'^{(m-1)}_k\rangle \}$ with nonzero 
eigenvalues $\{y^{(m)}_k\}$. 
They span the subspace $Q_m$. 
When all the eigenvalues $\{y^{(m)}_k\}$ are zero, the procedure ends because the eigenstates 
of $H$ reside in the subspace $P+Q_1+Q_2+\cdots+Q_{m-1}$. 
Here we note that we are not interested in any eigenstates that are decoupled from the states 
in the $P$ space. 
With the projection operators $Q_m$ and $Q_{m-1}$ we obtain, 
from Eq.(\ref{eq:q(m)_k-state-definition}), an expression written as 
\begin{align}
\label{eq:Qm-H-Qm-1}
 Q_mHQ_{m-1} = \sum_{k=1}^{d_m} \sqrt{y_k^{(m)}}
   |q_k^{(m)}\rangle \langle q_k'^{(m-1)}|.
\end{align}

\begin{figure}[h]
\begin{center}
\includegraphics[width=10cm]{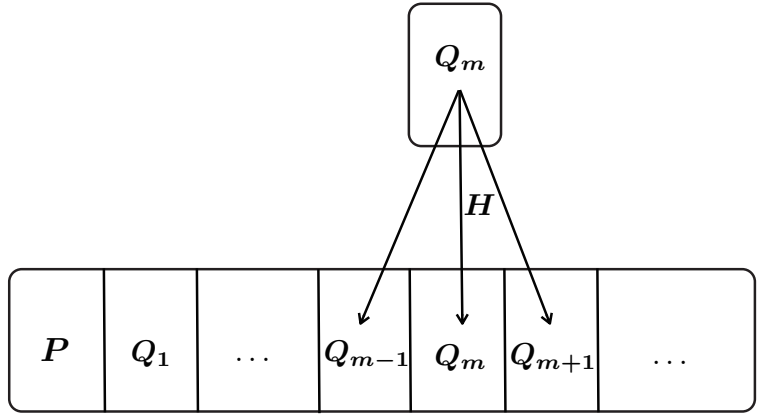}
\caption{The image $H(Q_m)$ by the mapping $H$ for $m\ge2$.
The $Q_{m-1}$, $Q_m$ and $Q_{m+1}$ are the subspaces of the $Q$ space which constitute the image $H(Q_m)$.
}

\label{fig : mapping-Qm}
\end{center}
\end{figure}

We conclude from the above discussion  that 
\begin{align}
\label{eq:PHQ_m-decoupling}
 PHQ_m&=Q_mHP=0\,\,\,\,(m \ge 2),\\
 \label{eq:QmHQm+k-decoupling}
  {Q}_m HQ_{m+k} &= {Q}_{m+k} HQ_{m}=0\,\,\,\,(k\ge 2)  
\end{align}
hold for the subspaces $\{P, Q_1, Q_2, \cdots, Q_m, \cdots\}$. 
This means that the given Hamiltonian $H$ is transformed to a block-tridiagonal matrix 
\begin{eqnarray}
\label{eq:H-decomposition}
H=\begin{pmatrix}
    PHP      & PHQ_1     &      0        &    0        & \cdots \cr
    Q_1 HP  & Q_1 HQ_1 & Q_1 HQ_2  &    0         & \cdots \cr
    0          & Q_2 HQ_1 & Q_2 HQ_2  &  Q_2HQ_3 & \cdots \cr
    0          &       0      & Q_3 HQ_2  &  Q_3HQ_3 & \cdots \cr
    \vdots   & \vdots    & \vdots      & \vdots     &\vdots 
\end{pmatrix} 
,
\end{eqnarray}
where each block matrix is at most $d$-dimensional. Thus the image $H(Q_m)$ by the mapping $H$ 
is a sum of adjacent subspaces $Q_{m-1}$, $Q_m$, and $Q_{m+1}$ as depicted in 
Fig.{\ref{fig : mapping-Qm}}. 
%

From Figs.\ref{fig : mapping-P} and \ref{fig : mapping-Qm} it is easy to see 
that the image of the mapping $H$ of the $P$ space becomes 
\begin{align}
\label{eq:mapping-H(P)}
    H(P) = P + Q_1 .
\end{align}
The image of the successive mapping is given by 
\begin{align}
\label{eq:mapping-H(P)-2}
    H^2(P) &= H(P + Q_1) \nonumber\\
                 &=  P + Q_1 + Q_2
\end{align}
and generally
\begin{align}
\label{eq:mapping-Hm(P)}
    H^m(P) = P + Q_1 + Q_2+\cdots +Q_m.
\end{align}

The above relations mean that the mapping  $H^m(P)$ generates an additional subspace $Q_{m}$.  
The sequence $\{P, H(P), \cdots , H^m(P) \} $ is called the Krylov subspaces \cite{GL96}.  
It may be clear that the subspaces $\{P, Q_1, \cdots , Q_{m}\} $ determine 
a unique block-tridiagonal form of $H$.  
In this sense the subspaces introduced in the present approach are essentially the same 
as those of Krylov.  
However, the basis states of each subspace $Q_k$ are ambiguous.
Determination of the basis states depends on the purpose, that is, what problem we want to solve 
after the block tridiagonalization of the Hamiltonian. 
We show, in the later sections, that the basis states introduced in the present study are useful 
for the formulation of the effective-interaction theory. 

%
%
\subsection{Expression of the $\widehat Q$ box in terms of the $\omega$ operator}

Here we define two operators
\begin{align}
\label{eq:eE}
       e(E) &= Q(E-H)Q , \\
 \label{eq:omega-E}
       \chi (E) &=  \frac{1}{e(E)}QHP \notag \\
                &=  \frac{1}{E-QHQ}QHP. 
\end{align}
In terms of $\chi (E)$, the $\widehat Q$ box in Eq.(\ref{eq:Q-box}) 
is expressed as 
\begin{align}
\label{eq:Q-box-new}
\widehat{Q}(E) = PHP+PHQ\chi (E)
\end{align}
and the solution $\omega$ in Eq.(\ref{eq:omega-representation}) to the decoupling 
equation (\ref{eq:decoupling-eq2}) is given by
\begin{align}
\label{eq:solution-omega}
\omega = \sum_{k=1}^d \chi (E_k) |\phi_k\rangle \langle \widetilde\phi_k|,
\end{align}
where $ |\phi_k\rangle $ and $\langle \widetilde\phi_k|$ have been defined 
in Eqs.(\ref{eq:eff-Hamiltonian-eigenvalue-eq}) and (\ref{eq:bi-orthogonal-state}).
Consequently calculating $\widehat{Q}(E)$ reduces to 
that of $\chi (E)$. 
When the $Q$ space is decomposed as in Eq.(\ref{eq:Q-decomposition}), 
also $\chi (E)$ is as 
\begin{align}
\label{eq:omega(E)-decomposition}
 \chi (E) &= \chi_{1}(E)+ \chi_{2}(E)+\cdots 
     + \chi_n(E) +\cdots,
\end{align}
where 
\begin{align}
\label{eq:omega_k(E)}
\chi_n(E) &=Q_n \ \chi (E)P.
\end{align}
Coupled equations for $\{ \chi_n (E) \}$ 
\begin{align}
\label{eq:Q1e(E)}
Q_1\,e(E)\{ \chi_1 (E)+\chi_2 (E)\} &= Q_1 HP,\\
\label{eq:Q2e(E)}
Q_2\,e(E)\{ \chi_1 (E)+\chi_2 (E)+\chi_3 (E)\} &= 0 ,\\
\vdots \cr
\label{eq:QNe(E)}
Q_n \,e(E)\{ \chi_{n-1} (E)+\chi_n (E) +\chi_{n+1}(E)\} &= 0 ,\\
\vdots \nonumber
\end{align}
are derived from Eq.(\ref{eq:omega-E}) using 
Eqs.(\ref{eq:QHP-2}), (\ref{eq:PHQ_m-decoupling}) and 
(\ref{eq:QmHQm+k-decoupling}). 
Since the $\widehat Q$ box is expressed as 
\begin{align}
\label{eq:Q-box-new-2}
\widehat{Q}(E) &= PHP+PHQ_{1} \chi_{1}(E)
\end{align}
by using Eq.(\ref{eq:QHP-2}), calculating the $\widehat Q$ box 
reduces to that of $\chi_{1}(E)$.
%
%
\subsection{Expansion in terms of continued fraction}
We show that the $\widehat Q$ box is expanded by 
a continued fraction \cite{JT80} 
of small-dimensional matrices by solving Eqs.(\ref{eq:Q1e(E)})\,--\,(\ref{eq:QNe(E)}). 
We assume $\chi_m(E)=0$ for $m\ge 2$ , then we have
\begin{align}
\label{eq:omega-1(E)}
\chi_{1}(E) &= \frac{1}{e_{1}(E)}Q_{1}HP
\end{align}
from Eq.(\ref{eq:Q1e(E)}), where
\begin{align}
\label{eq:e-1(E)}
e_{1}(E) &= Q_{1}(E-H)Q_{1}.
\end{align}
Hereafter we use the notation
\begin{align}
\label{eq:e-m(E)}
e_{m}(E) &= Q_{m}(E-H)Q_{m}.
\end{align}
The solution (\ref{eq:omega-1(E)}) gives the $\widehat Q$ box 
in the first approximation as 
\begin{align}
\label{eq:Q-box(1)}
\widehat{Q}^{(1)}(E) &= PHP+PHQ_{1}\frac{1}{e_{1}(E)}Q_{1}HP. 
\end{align}

Next we have 
\begin{align}
\label{eq:omega-2(E)}
\chi_{2}(E) &= \frac{1}{e_{2}(E)}Q_{2}HQ_{1}\chi_{1}(E)
\end{align}
from Eq.(\ref{eq:Q2e(E)}) by assuming $\chi_{m}(E)=0$ for $m\ge 3$. 
Substituing this into Eq.(\ref{eq:Q1e(E)}) leads to 
\begin{align}
\label{eq:omega-1(E)-new}
\chi_{1}(E) &=
\dfrac{1}{e_{1}(E)-Q_{1}HQ_{2}\dfrac{1}{e_{2}(E)}Q_{2}HQ_{1}}Q_{1}HP,
\end{align}
and then the $\widehat Q$ box is given as 
\begin{align}
\label{eq:Q-box(2)}
\widehat{Q}^{(2)}(E) &= PHP+PHQ_{1}\frac{1}{e_{1}(E)-Q_{1}HQ_{2}
\dfrac{1}{e_{2}(E)}Q_{2}HQ_{1}}Q_{1}HP
\end{align}
in the second approximation. Repeating similar manipulations, we finally have a general form 
\begin{align}
\label{eq:Q-box-continued-frac}
\widehat{Q}(E) 
    &= PHP+PHQ_{1}\dfrac{1}{e_{1}-H_{12}\dfrac{1}{e_{2}-H_{23}
         \dfrac{1}{e_3-H_{34}\dfrac{1}{e_4 - \cdots}
         H_{43}}H_{32}}H_{21}}Q_1HP
\end{align}
with $e_{m}=e_{m}(E)$ and
\begin{align}
\label{eq:Hij}
H_{ij} &= Q_i HQ_j .
\end{align}

Here we consider a case in which the $Q$ space for a system of interest 
is well described by finite number of subspaces. 
We denote the maximum of 
$n$ by $N$ in Eq.(\ref{eq:Q-decomposition}). 
We introduce $\{{\widetilde e}_n(E)\}$ 
given through a descending recurrence relation starting from $n=N$ as
\begin{align}
\label{eq:tilde-e-(n-1)}
\widetilde{e}_{n-1}(E)
     &= {e}_{n-1}(E)-H_{n-1,n}\frac{1}{\widetilde{e}_{n}(E)}H_{n,n-1} , 
\end{align}
where we define 
\begin{align}
\label{eq:tilde-e-(N)}
\widetilde{e}_{N}(E) &= Q_N (E-H) Q_N.
\end{align}
From Eq.(\ref{eq:tilde-e-(n-1)}) we have a sequence 
$\widetilde{e}_{N-1}(E)$, $\widetilde{e}_{N-2}(E)$, $\cdots$, and $\widetilde{e}_{1}(E)$. 
Then the $\widehat Q$ box is expressed as 
\begin{align}
\label{eq:Q-box-tilde-e-(1)}
\widehat{Q}(E) = PHP+PHQ_1\frac{1}{\widetilde{e}_{1}(E)}Q_1 HP.
\end{align}
%
\begin{figure}[h]
   \begin{center} 
      \includegraphics[width=10cm,keepaspectratio]{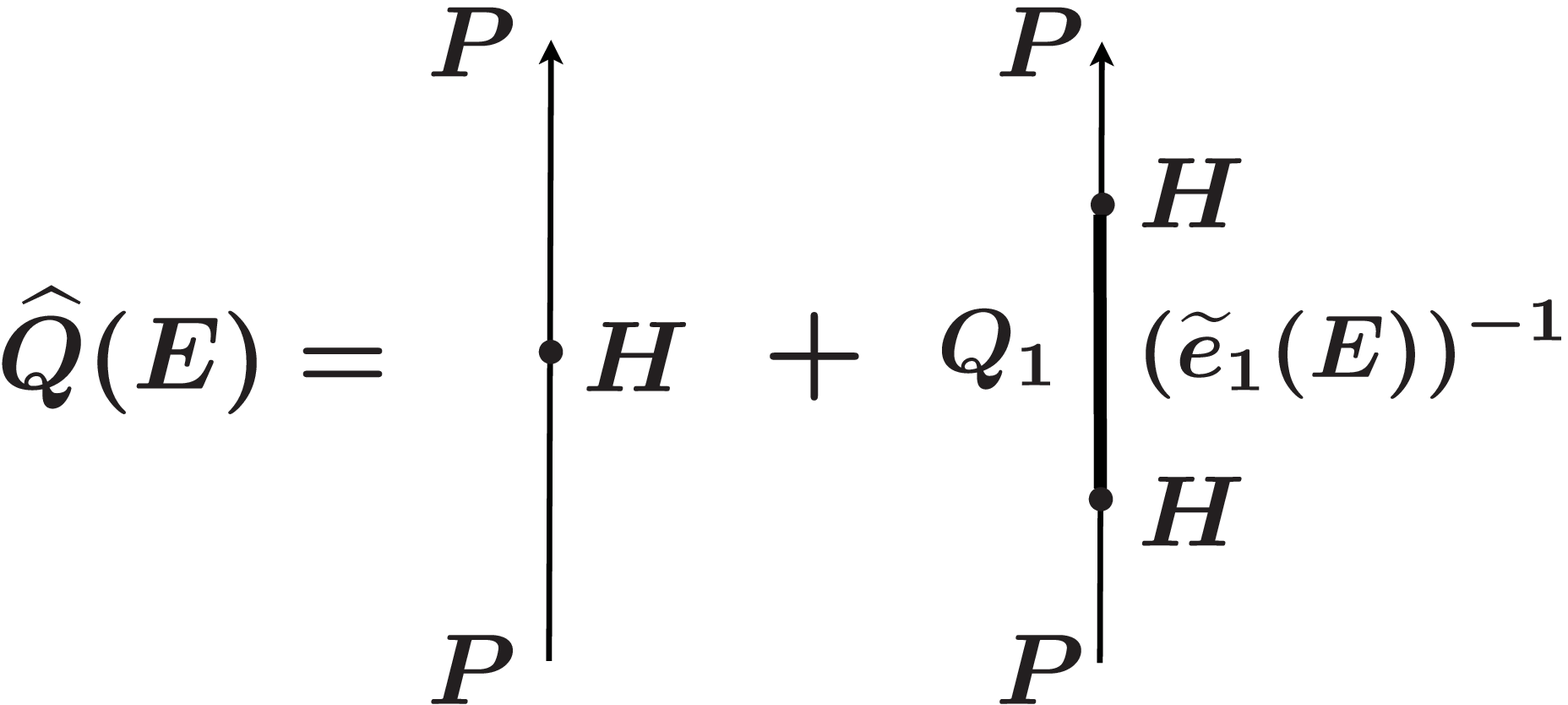} 
   \caption{Diagrammatical expression of the $\widehat Q$ box
         in terms of the renormalized propagator $(\widetilde{e}_1(E))^{-1}$ 
which is composed of the continued fraction. 
The $H$ denotes the Hamiltonian. The $P$ and $Q_1$ are the projection operators onto the model 
space and the $Q_1$ space, respectively, where the $Q_1$ space is the $Q$-space part of the image $H(P)$.
The thick like expresses the propagation of $Q_1$-space states with the propagator 
$(\widetilde{e}_1(E))^{-1}$. }
  \label{fig : perturbation-Q(E)-continued-fraction} 
  \end{center}
\end{figure}  
Diagrammatical expression of $\widehat Q(E)$ is shown in 
Fig.\ref{fig : perturbation-Q(E)-continued-fraction}.
It is a remarkable fact that the above result for the $\widehat Q$ box 
indicates the existence of the  renormalized inverse propagator 
$\widetilde{e}_{1}(E)$ such that the $\widehat Q$ box can be represented 
by a sum of only two terms, namely, the unperturbed part and the 
second-order term.

If the dimension of the $Q$ space is finite, the number of the subspaces 
$\{Q_m\}$ 
is also finite and the $\widehat Q$ box given in Eq.(\ref{eq:Q-box-tilde-e-(1)}) 
is exact.
On the other hand, if the dimension of the $Q$ space is infinite, 
the number of the subspaces $\{Q_m\}$ is, in general, infinite.
For this case we introduce a truncation of the $Q$ space.
We consider a finite-dimensional subspace $Q_1 + Q_2 + \cdots + Q_N$, where
the subspaces $\{Q_m, 1 \leq m \leq N\}$ lead to a block-tridiagonal form of 
$H$ as in Eq.(\ref{eq:H-decomposition}).
The operator $\widetilde e_1(E)$ that is determined through the recurrence relation 
in Eq.(\ref{eq:tilde-e-(n-1)}) starting with $n=N$ is a function of $N$ and we write it as 
$\widetilde e_1^{(N)}(E)$. 
If $\widetilde e_1^{(N)}(E)$ converges as $N$ tends to infinity, we can write 
the $\widehat Q$ box as 
\begin{align}
\label{eq:Q-box-tilde-e-(1)-infinite}
\widehat{Q}(E) &= PHP+PHQ_1\frac{1}{\widetilde{e}_{1}^{(\infty)}(E)}Q_1 HP,
\end{align}
where 
\begin{align}
\label{eq:tilde-e-(1)-infinite}
\widetilde{e}_{1}^{(\infty)}(E) 
     = \lim_{N\rightarrow \infty} \widetilde{e}_{1}^{(N)}(E).
\end{align}

We discuss the meaning of Eq.(\ref{eq:Q-box-tilde-e-(1)-infinite}) in more detail. 
We consider an application of the present formalism to the calculation of the 
effective interaction between two valence nucleons outside a core, such as $^{16}$O. 
Many of the numerical calculations have shown that the second-order diagrams make 
dominant contributions \cite{SKS83, ACCGKLP12} and the third-and-higher-order terms 
are less important. 
It should be pointed out that, in many of such calculations, the experimental 
single-particle (s.p.) energies have been employed. 
As shown in Eqs.(\ref{eq:Q-box-tilde-e-(1)}) 
and (\ref{eq:Q-box-tilde-e-(1)-infinite}) the $\widehat Q$ box can be expressed finally 
as the second-order diagrams with the unchanged (not renormalized) vertex $PHQ_1$ (=$PHQ$) 
and the renormalized inverse propagator $\widetilde{e}_1(E)$. 
This fact means that, if we use a proper 
$\widetilde{e}_1(E)$, the exact $\widehat Q$ box can be given by the second-order term. 
There is a possibility that $\widetilde{e}_1(E)$ can be replaced approximately 
with the energy denominator determined from the experimental s.p. energies. 
We mention that the expression 
of the $\widehat Q$ box in Eqs.(\ref{eq:Q-box-tilde-e-(1)}) or (\ref{eq:Q-box-tilde-e-(1)-infinite}) 
would give an explanation for the reason why the second-order diagrams make dominant 
contributions and lead to fairly good agreement with the experimental spectra.
%
%
\subsection{Expansion in terms of renormalized vertices and propagators}

We here consider a method of calculation by an ascending 
recurrence relation for $\{\chi_n(E)\}$ 
and derive another solution for the $\widehat Q$ box.
By using Eqs.(\ref{eq:eE}) and (\ref{eq:Hij}), the coupled equations
Eqs.(\ref{eq:Q1e(E)})\,--\,(\ref{eq:QNe(E)}) for the operators
 $\{\chi_n(E)\}$ are written as
\begin{align}
\label{eq:e1omega1}
e_{1}(E) \chi_1 (E) &= H_{10}+H_{12}\chi_{2}(E),\\
\label{eq:e2omega2}
e_{2}(E) \chi_2 (E) &= H_{21}\chi_1 (E)+H_{23}\chi_{3}(E),\\
\vdots \cr
\label{eq:ekomegak}
e_{n}(E) \chi_n (E) 
      &= H_{n,n-1}\chi_{n-1} (E)+H_{n,n+1}\chi_{n+1}(E),\\
\vdots \nonumber
\end{align}
with 
\begin{align}
\label{eq:H10}
H_{10} &= Q_{1}HP.
\end{align}
Equations (\ref{eq:ekomegak}) is a linear relation of three operators 
$\chi_{n-1}, \chi_n$, and $\chi_{n+1}$, which can be cast into those of two operators 
as follows: 
First we rewrite (\ref{eq:e1omega1}) as 
\begin{align}
\label{eq:omega_1}
\chi_{1}(E) &= \alpha_{1}(E)+\beta_{1}(E)\chi_{2}(E)
\end{align}
with 
\begin{align}
\label{eq:alpha_1}
\alpha_{1}(E) &= \frac{1}{e_{1}(E)}H_{10},\\
\label{eq:beta_1}
\beta_{1}(E) &= \frac{1}{e_{1}(E)}H_{12}.
\end{align}
By substituting this into Eq.(\ref{eq:e2omega2}), 
$\chi_2 (E)$ is expressed as linear with $\chi_3(E)$, 
\begin{align}
\label{eq:omega_2}
\chi_{2}(E) &= \alpha_{2}(E)+\beta_{2}(E)\chi_{3}(E), 
\end{align}
where 
\begin{align}
\label{eq:alpha_2}
\alpha_{2}(E) &= \frac{1}{e_{2}(E)-H_{21}\dfrac{1}{e_{1}(E)}H_{12}}
                         H_{21}\frac{1}{e_{1}(E)}H_{10}\cr
      &= \frac{1}{e_{2}(E)-H_{21}\beta_{1}(E)}  H_{21} \alpha_{1}(E),
\\
\label{eq:beta_2}
\beta_{2}(E)&=\frac{1}{e_{2}(E)-H_{21}\dfrac{1}{e_{1}(E)}H_{12}}H_{23}\cr
      &=\frac{1}{e_{2}(E)-H_{21}\beta_{1}(E)}  H_{23}.
\end{align}
In general, we define the operators $\alpha_n(E)$ and $\beta_n(E)$ that obey the following 
ascending recurrence relations, 
\begin{align}
\label{eq:alpha_n}
\alpha_{n}(E) &= \frac{1}{e_{n}(E)-H_{n,n-1}\beta_{n-1}(E)}
                         H_{n,n-1}\alpha_{n-1}(E),\\
\label{eq:beta_n}
\beta_{n}(E) &= \frac{1}{e_{n}(E)-H_{n,n-1}\beta_{n-1}(E)} H_{n,n+1}.
\end{align}
We then have a linear relation 
\begin{align}
\label{eq:omega_n}
\chi_{n}(E) &= \alpha_{n}(E)+\beta_{n}(E)\chi_{n+1}(E).
\end{align}
Equations (\ref{eq:alpha_n}) and (\ref{eq:beta_n}) determine 
$\{ \alpha_{n}(E),\beta_{n}(E), n=1,2,\cdots\}$ with the initial values 
$\alpha_1(E)$ and $\beta_1(E)$ in Eqs.(\ref{eq:alpha_1}) and (\ref{eq:beta_1}),
respectively. We finally have a solution for $\chi_{1}(E)$ as 
\begin{align}
\label{eq:omega_1_solution}
\chi_{1}(E) 
     &= \alpha_{1}(E)+\beta_{1}(E)\alpha_{2}(E)
         +\cdots
         +\beta_{1}(E)\beta_{2}(E)\cdots\beta_{n-1}(E)\alpha_{n}(E) + \cdots \cr
     &= \sum_{k=1}^{\infty}{\Big\{ }\prod_{m=1}^{k-1}\beta_{m}(E){\Big\} }\alpha_{k}(E).
\end{align}
Consequently the $\widehat Q$ box is given by 
\begin{align}
\label{eq:Q_box_solution}
\widehat{Q}(E) &= PHP  +PHQ_{1}\left [ \sum_{k=1}^{\infty}
                           {\Big\{ }\prod_{m=1}^{k-1}\beta_{m}(E){\Big\} } \alpha_{k}(E)\right ].
\end{align}

\begin{figure}[h]
   \begin{center} 
\includegraphics[width=10cm,keepaspectratio]{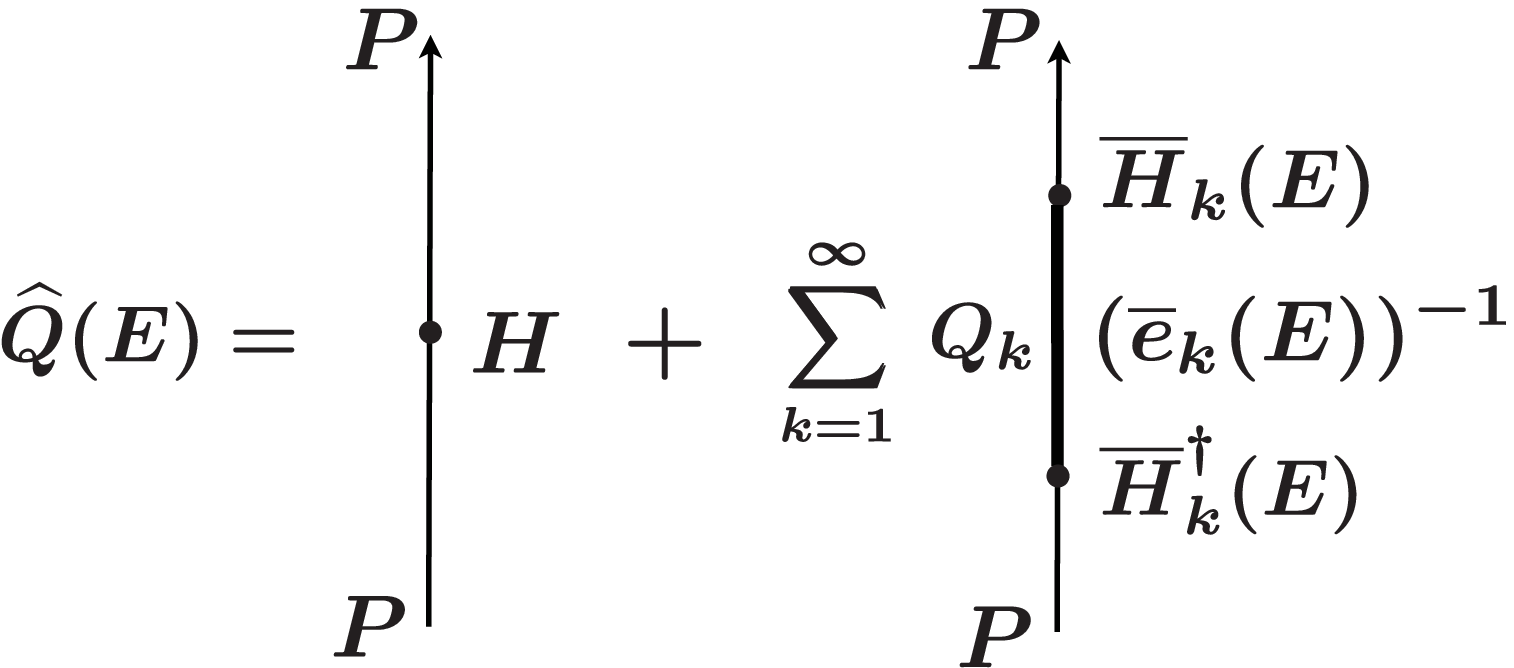}     
   \caption{Diagrammatical expression of 
the $\widehat Q$ box in terms of the renormalized vertices $\overline{H}_{k}(E)$ and the 
propagators $(\overline{e}_k(E))^{-1}$. Other notations are the same as in Fig.
\ref{fig : perturbation-Q(E)-continued-fraction}.
}
  \label{fig : perturbation-Q(E)-renormalized-propagator} 
  \end{center}
\end{figure}

In order to rewrite $\{\alpha_{n}(E)\}$, $\{\beta_{n}(E)\}$, and the $\widehat Q$ box 
in terms of $\{e_{i}(E)\}$ and $\{H_{ij}\}$, 
we introduce another inverse propagator $\overline{e}_{m}(E)$ 
defined through the following recurrence relation 
\begin{align}
\label{eq:inverse_propagator}
\overline{e}_{m}(E) &= e_{m}(E)-H_{m,m-1}\frac{1}{\overline{e}_{m-1}(E)}H_{m-1,m}
\end{align}
with the initial value 
\begin{align}
\label{eq:inverse_propagator_1}
\overline{e}_{1}(E) &= e_{1}(E)\cr
                  &= Q_{1}(E-H)Q_{1}.
\end{align}
We note that $\overline{e}_{m}(E)$ in Eq.(\ref{eq:inverse_propagator}) obeys 
an ascending recurrence relation, which differs from $\widetilde{e}_m(E)$ in 
Eq.(\ref{eq:tilde-e-(n-1)}). 
In terms of $\{\overline{e}_{m}(E)\}$, 
the operators $\{\alpha_{n}(E)\}$ and $\{\beta_{n}(E)\}$ are written as 
\begin{align}
\label{eq:alpha_inverse_propagator_1}
\alpha_{1}(E) &= \frac{1}{\overline{e}_{1}(E) }H_{10} ,\\
\label{eq:alpha_inverse_propagator_2}
 \alpha_{2}(E) 
     &= \frac{1}{\overline{e}_{2}(E) }H_{21}\frac{1}{\overline{e}_{1}(E) }H_{10} ,\\
   \vdots \cr
\label{eq:alpha_inverse_propagator_k}
 \alpha_{n}(E) 
    &= \frac{1}{\overline{e}_{n}(E) }H_{n,n-1}\frac{1}{\overline{e}_{n-1}(E)}
                              H_{n-1,n-2}\cdots H_{21}\frac{1}{\overline{e}_{1}(E) }H_{10},\\
     \vdots \cr
\label{eq:beta_inverse_propagator_1}
\beta_{1}(E) &= \frac{1}{\overline{e}_{1}(E) }H_{12} ,\\
\label{eq:beta_inverse_propagator_2}
 \beta_{2}(E) &= \frac{1}{\overline{e}_{2}(E) }H_{23} ,\\
   \vdots \cr
\label{eq:beta_inverse_propagator_k}
 \beta_{n}(E) &= \frac{1}{\overline{e}_{n}(E) }H_{n,n+1},\\
 \vdots \,\,\,. \cr\nonumber
\end{align}
Then the $\widehat Q$ box in Eq.(\ref{eq:Q_box_solution}) is expressed explicitly as 
\begin{eqnarray}
\label{eq:Q_box_solution_inverse_propagator}
\widehat{Q}(E) 
     &=& PHP  +H_{01}\frac{1}{\overline{e}_{1}(E) }H_{10}
             +H_{01}\frac{1}{\overline{e}_{1}(E) }H_{12}
               \frac{1}{\overline{e}_{2}(E) }H_{21}\frac{1}{\overline{e}_{1}(E) }H_{10}
               +\cdots\cr
     && \hspace{-2mm}+H_{01}\frac{1}{\overline{e}_{1}(E) }H_{12}\cdots 
            H_{n-1,n}\frac{1}{\overline{e}_{n}(E) }H_{n,n-1}\cdots
            H_{21}\frac{1}{\overline{e}_{1}(E) }H_{10}+\cdots.
\end{eqnarray}
A simpler expression of the $\widehat Q$ box can be obtained by utilizing 
$\{ \overline{H}_{k}(E)\}$ defined through 
\begin{align}
\label{eq:overline_H_k+1}
\overline{H}_{k}(E) 
        &= H_{01}\frac{1}{\overline{e}_{1}(E) }H_{12}\frac{1}{\overline{e}_{2}(E) }H_{23}
                \cdots \frac{1}{\overline{e}_{k-1}(E) }H_{k-1,k}\cr
        &= \overline{H}_{k-1}(E) \frac{1}{\overline{e}_{k-1}(E) }H_{k-1,k}      
\end{align}
with the initial value 
\begin{align}
\label{eq:overline_H_1}
\overline{H}_{1}(E) = PHQ_{1}.
\end{align}
The $\overline{H}_{k}(E)$ interconnecting the $P$ and $Q_k$ spaces is 
a $d$$\times$$d_k$ matrix. 
The $\widehat{Q}(E)$ in Eq.(\ref{eq:Q_box_solution_inverse_propagator}) 
is further reduced to 
\begin{align}
\label{eq:Q_box_solution_inverse_propagator_final}
\widehat{Q}(E) 
   &= PHP+\overline{H}_{1}(E)\frac{1}{\overline{e}_{1}(E)}\overline{H}^{\dagger}_{1}(E)
     +\cdots+\overline{H}_{n}(E)\frac{1}{\overline{e}_{n}(E)}\overline{H}^{\dagger}_{n}(E)
     +\cdots\cr
   &= PHP+\sum_{k=1}^{\infty}\overline{H}_{k}(E)
                   \frac{1}{\overline{e}_{k}(E)}\overline{H}^{\dagger}_{k}(E) .
\end{align}
This expression can be interpreted as that the $\widehat Q$ box is given by a sum up 
to second order in the usual perturbation theory as schematically depicted 
in Fig.\ref{fig : perturbation-Q(E)-renormalized-propagator} in terms of the renormalized 
inverse propagators $\{\overline{e}_k(E)\}$ and the renormalized vertices 
$\{\overline H_k(E)\}$. 
%
%
Equation (\ref{eq:Q_box_solution_inverse_propagator_final}) shows clearly that there exist 
the renormalized inverse propagators $\{\bar e_k(E)\}$ and the renormalized vertices 
$\{\bar H_k(E)\}$ such that the $\widehat Q$ box can be represented by a second-order-perturbation 
form which is the lowest-order interaction terms. 

If a system with a Hamiltonian $H$ can be well described in a finite-dimensional space,
the $\widehat Q$ box in Eq.(\ref{eq:Q_box_solution_inverse_propagator}) is given by 
a sum of finite number of terms and should coincide with that in Eq.(\ref{eq:Q-box-tilde-e-(1)}). 
Compairing two solutions for the $\widehat Q$ box, we have an expression of the renormalized 
propagator  $\{\widetilde e_1(E)\}^{-1}$ as 
\begin{eqnarray}
\label{eq:inverse_tilde_e1_expansion}
\frac{1}{\widetilde e_1(E)} 
     &=& \frac{1}{\overline e_1(E)}  
             +\frac{1}{\overline{e}_{1}(E) }H_{12}
               \frac{1}{\overline{e}_{2}(E) }H_{21}\frac{1}{\overline{e}_{1}(E) }
               +\cdots\cr
     && \hspace{-2mm}+\frac{1}{\overline{e}_{1}(E) }H_{12}\cdots 
            H_{N-1,N}\frac{1}{\overline{e}_{N}(E) }H_{N,N-1}\cdots
            H_{21}\frac{1}{\overline{e}_{1}(E)},
\end{eqnarray}
where $N$ is the number of the subspaces $\{Q_k\}$.
The above $\widetilde e_1(E)$ can be a solution to the recursive equation (\ref{eq:tilde-e-(n-1)}) 
and gives  an expansion formula in terms of $\{H_{k-1, k}\}$, $\{H_{k, k-1}\}$ and 
$\{\overline{e}_{k}(E)\}$ which are defined with the subspaces $\{Q_k\}$. 
Recall that the calculation of the $\widehat Q$ box is reduced to that of $\widetilde e_1(E)$ as 
in Eq.(\ref{eq:Q-box-tilde-e-(1)}). 
The expression of $\{\widetilde e_1(E)\}^{-1}$ in Eq.(\ref{eq:inverse_tilde_e1_expansion}) 
makes it clear how the subspaces $\{Q_k\}$ contribute to $\widetilde e_1(E)$ and, 
equivalently, to the $\widehat Q$ box. 
Therefore, when we consider introducing an approximation in a practical problem, 
Eq.(\ref{eq:inverse_tilde_e1_expansion}) would provide us with a basic formula for  
$\{\widetilde e_1(E)\}^{-1}$.
%
%
\section{Recursive solution for the $\chi (E)$ operator}
We here discuss how to calculate the operator $\chi_n(E)$ in Eq.(\ref{eq:omega_k(E)}) which are necessary
for obtaining a true eigenstate, namely, $|\Phi_k\rangle$ with the eigenvalue $E=E_k$.
The basic equations for determining $\{\chi_n(E)\}$ have been given 
in subsections III B and III D. 
In the similarity-transformation theory for the effective interaction, the relationship 
between $|\Phi_{k}\rangle$ and the model-space eigenstate $|\phi_{k}\rangle$ is 
\begin{align}
\label{eq:eigenstate-Phi-phi}
|\Phi_{k}\rangle   &= e^{\omega} |\phi_{k}\rangle\cr
                              &= |\phi_{k}\rangle +\omega |\phi_{k}\rangle .
\end{align}
Using Eq.(\ref{eq:solution-omega}) for $\omega$ 
in terms of $\chi (E_{k})$, $|\Phi_{k}\rangle $ is also expressed as 
\begin{align}
\label{eq:eigenstate-Phi-phi-a}
|\Phi_{k}\rangle  = |\phi_{k}\rangle +\chi (E_k) |\phi_{k}\rangle .
\end{align}
Therefore, if we want to obtain $|\Phi_{k}\rangle$, we have to solve $\chi (E_k)$.
We decompose $\chi (E_k)$ into $\{\chi_n (E_k)\}$ as 
in Eq.(\ref{eq:omega(E)-decomposition}).
The sequence $\{\chi_1 (E_k),\, \chi_2 (E_k),\cdots \}$ obeys 
Eqs.(\ref{eq:e1omega1})--(\ref{eq:ekomegak}).
From Eq.(\ref{eq:ekomegak}) the following recurrence relation is obtained for 
$\{\chi_n (E_k)\}$;
\begin{align}
\label{eq:eq:omegan-en}
\chi_{n+1}(E_k)  
     =K_{n+1,n}\{e_n(E_k)\chi_n(E_k)  - H_{n,n-1}\chi_{n-1}(E_k)\}
     \hspace{0.5cm}(n \geq 2),
\end{align}
where $K_{n+1,n}$ is defined as
\begin{align}
\label{eq:define-K}
K_{n+1,n}  
     =\sum_{k=1}^{d_{n+1}}\frac{1}{\sqrt{y_k^{(n+1)}}}
                       |q_{k}^{(n+1)}\rangle \langle  {q'}_{k}^{(n )}|.
\end{align}
It is easy to see, using Eqs.(\ref{eq:Qm-H-Qm-1}) and (\ref{eq:Hij}) for $H_{n,n+1}$,
\begin{align}
\label{eq:KH=Q}
K_{n+1,n} \cdot H_{n,n+1} = Q_{n+1},
\end{align}
from which Eq.(\ref{eq:eq:omegan-en}) is derived.
For the calculation of $\{\chi_n(E_k)\}$ with $n \geq 3$, 
$\chi_1(E_k)$ and $\chi_2(E_k)$ are necessary as initial values. 
In this stage we suppose that $\widehat Q(E_k)$ is given beforehand and
use Eq.(\ref{eq:Q-box-new-2}) to obtain 
\begin{align}
\label{eq:omega1-ek}
\chi_1(E_k) = K_{10} (\widehat Q (E_k) -PHP)
\end{align}
with 
\begin{align}
\label{eq:K10}
 K_{10} = \sum_{k=1}^{d_1} \frac{1}{\sqrt{y_k^{(1)}}}
                       |q_{k}^{(1)}\rangle \langle  p_{k}|,
 \end{align}
where $\langle  p_{k}|$ and $ |q_{k}^{(1)}\rangle$ are given 
in Eqs.(\ref{eq:Yp-operator-eigenvalue-eq}) and (\ref{eq:q_k-state-definition}), respectively.
In a similar manner, the operator $\chi_2(E_k)$ is solved, 
using Eq.(\ref{eq:e1omega1}), as
\begin{align}
\label{eq:omega2}
 \chi_2(E_k) = K_{21}\cdot \{e_1(E_k)\chi_1(E_k) - H_{1,0}\}
 \end{align}
with
\begin{align}
\label{eq:K21}
 K_{21} = \sum_{k=1}^{d_2} \frac{1}{\sqrt{y_k^{(2)}}}
                       |q_{k}^{(2)}\rangle \langle  {q'}_{k}^{(1)}|,
 \end{align}
where $\langle  {q'}_{k}^{(1)}|$ and $|q_{k}^{(2)}\rangle$ are given 
in Eqs.(\ref{eq:Yq1-operator-eigenvalue-eq}) and (\ref{eq:q(2)_k-state-definition}), respectively.
Substituting $\chi_1(E_k)$ and $\chi_2(E_k)$, 
the sequence $\chi_3(E_k)$, $\chi_4(E_k)$, $\cdots$ are obtained from the recurrence 
relation in Eq.(\ref{eq:eq:omegan-en}).

The eigenstate $|\Phi_k \rangle$ of $H$ with the eigenvalue $E_k$ is
finally given by
\begin{align}
\label{eq:eigenstate-Psi-k}
 |\Phi_k \rangle = |\phi_k \rangle 
                              + \sum_{n} \chi_n(E_k) |\phi_k \rangle .
 \end{align}
The usual normalization in the effective-interaction theory is 
$\langle \phi_k|\phi_{k'} \rangle =\delta_{k,k'}$.
Therefore, the normalized true eigenstate denoted by $ |\Psi_k \rangle$
is given by 
\begin{align}
\label{eq:true-eigenstate-Psi-k}
       |\Psi_k \rangle =\frac{1}{N_k} |\Phi_k \rangle  ,
 \end{align}
where the normalization factor $N_k$ is
\begin{align}
\label{eq:normaization-factor-Nk}
N_k = \sqrt{1+ 
 \sum_{n}
      \langle \phi_k | \chi_n^{\dagger}(E_k)\chi_n(E_k) |\phi_k \rangle} .
 \end{align}
%
\section{The $\widehat Z$-box method and effective Hamiltonian}
\setcounter{equation}{0}
%
The $\widehat Z$ box has been defined in the previous paper \cite{SOKF11} as 
\begin{align}
\label{eq:Z-box-definition}
\widehat{Z}(E)
       & =  \frac{1}{1-\widehat{Q}_{1}(E)}
       {\big [}\widehat{Q}(E)-E\widehat{Q}_{1}(E){\big ]}
\end{align}
with 
\begin{align}
\label{eq:Q-box-differential-1}
\widehat{Q}_{1}(E) & =  \frac{d\widehat{Q}(E)}{dE}\nonumber\\
               &=-PHQ\frac{1}{(E-QHQ)^2}QHP,
\end{align}
in order to overcome some defects that inevitably accompany the 
$\widehat Q$-box approach. 
The $\widehat Z$ box has the following properties: 
\begin{enumerate}
\item[(i)]
The operator
\begin{align}
\label{eq:Heff-Z-box}
H_{\rm eff}
     &= \sum_{k=1}^{d}\widehat{Z}(E_k)|\phi_{k}\rangle\langle\widetilde{\phi}_k|,
\end{align}
which is obtained by replacing $\widehat{Q}(E)$ in Eq.(\ref{eq:tilde-H-representation-2}) 
with $\widehat{Z}(E)$, can be an effective Hamiltonian if $\{ E_k,k=1,2,\cdots,d\}$ 
are the eigenvalues of $H$. 
Therefore, $\widehat{Z}(E_k)$ satisfies the self-consistent equation
\begin{align}
\label{eq:Z-box-eigenvalue-eq}
 \widehat Z(E_k)|\phi_{k}\rangle = E_k|\phi_{k}\rangle .
\end{align}
\item[(ii)] 
The derivative of $\widehat{Z}(E)$ is given by 
\begin{align}
\label{eq:Z-box-differential}
\frac{d\widehat{Z}(E)}{dE} & =  
\frac{2}{1-\widehat{Q}_{1}(E)}\widehat{Q}_{2}(E)[\widehat{Z}(E)-EP]
\end{align}
with 
\begin{align}
\label{eq:Q-box-differential-2}
\widehat{Q}_{2}(E) &=
      \frac{1}{2!}\frac{d^{2}\widehat{Q}(E)}{dE^{2}}\nonumber\\
               &=PHQ\frac{1}{(E-QHQ)^3}QHP.
\end{align}
Then 
\begin{align}
\label{eq:Z-box-differential-1-at-eigenvalues}
\frac{d\widehat{Z}(E)}{dE}{\Big |}_{E=E_k}|\phi_k\rangle=0
\end{align}
holds for the eigenvalue $E_k$ and the corresponding eigenstate $|\phi_k\rangle$ 
of $H_{\rm eff}$. 
\item[(iii)] 
For the eigenvalue $\varepsilon_q$ of $QHQ$ determined by 
Eq.(\ref{eq:QHQ-eigenvalue-eq}), 
$\widehat{Z}(\varepsilon_q)$ satisfies the self-consistent equation
\begin{align}
\label{eq:Z-eigenvalue-eq-hole}
 \widehat{Z}(\varepsilon_q)|\mu_q\rangle&=\varepsilon_q|\mu_q\rangle.
\end{align}
Here we note that $|\mu_q\rangle$ belongs to the $P$ space. 
\item[(iv)] 
 Contrary to Eq.(\ref{eq:Z-box-differential-1-at-eigenvalues}),
\begin{align}
\label{eq:Z-box-differential-1-at-poles}
\frac{d\widehat{Z}(E)}{dE}{\Big |}_{E=\varepsilon_q}|\mu_q\rangle
     &= 2|\mu_q\rangle
\end{align}
holds for the derivative of $\widehat{Z}(E)$ at $E=\varepsilon_q$. 
\end{enumerate}

These properties lead to the conclusions: 
$\widehat{Z}(E)$ is finite and differentiable even at $E=\varepsilon_q$, 
a pole of $\widehat{Q}(E)$. Although $E=\varepsilon_q$ is also a solution
of the self-consistent equation for $\widehat{Z}(E)$, it can be easily discriminated 
from true eigenvalues $\{E_k\}$ of $H$ with the aid of their derivatives 
in Eqs.(\ref{eq:Z-box-differential-1-at-eigenvalues}) and 
(\ref{eq:Z-box-differential-1-at-poles}). The $\widehat Z$-box method has been applied recently 
to a realistic calculation of the effective interaction by Coraggio et al.\cite{CCGIK12}. 

In order to calculate the $\widehat Z$ box
we need the first and second derivatives of the $\widehat Q$ box.
These derivatives can be calculated analytically and are derived 
in Appendices A and B corresponding to two expressions of the $\widehat Q$ box 
given in Subsections III C and III D, respectively.
%
%
\section{Model calculation}
\subsection{Graphical method for eigenvalues of $H$}

We shall solve the eigenvalue problem for the Hamiltonian $H$ in the framework of 
the $\widehat Z$-box theory.
We note that the $\widehat Z$ box is a $d$-dimensional operator acting in the $P$
space and has $d$ eigenvalues. 
We have assumed that the operator 
$\widehat Z(E)$ for an arbitrary energy variable $E$ has $d$ different eigenvalues.
In the present calculation we do not discuss the case that $\widehat Z(E)$ has 
some degenerate eigenvalues. 
The eigenvalues of $\widehat Z(E)$ are functions of $E$.
We write the eigenvalue equation for $\widehat Z(E)$ as
\begin{align}
\label{eq:Z(E)-eigenvalue-eq}
 \widehat Z(E) |\zeta_k\rangle = F_k(E)|\zeta_k\rangle ,\,\,\,k=1,2,\cdots, d.
\end{align}
The above eigenvalue equation defines $d$ functions $\{F_k(E), k=1,2,\cdots, d\}$.
We label $\{F_k(E)\}$ in order of energy as $F_1(E) < F_2(E) < \cdots <F_d(E)$.
From Eq.(\ref{eq:Z-box-eigenvalue-eq}) we see that the solutions for the eigenvalues 
of $H$ can be obtained by solving 
\begin{align}
\label{eq:Fk(E)=E}
  F_k(E) = E.
\end{align}

As shown in the previous section, Eq.(\ref{eq:Z-box-eigenvalue-eq})  
has two kinds of solutions, namely, $E=E_i$ and $E=\varepsilon_j$, where $E_i$ 
and $\varepsilon_j$ are the eigenvalues of $H$ and $QHQ$, respectively.
We distinguish the eigenvalues $\{\varepsilon_j \}$ from $\{E_i\}$ according to the 
condition that the energy derivative $d\widehat Z /dE$ takes different values for 
$E=E_i$ and $E=\varepsilon_j$.
We define functions $\{F'_k(E), k=1,2,\cdots, d\}$ as 
\begin{align}
\label{eq:define-derivative-Fk(E)}
  F'_k(E) = \Big\langle \zeta_k \Big|\,\,\frac{d\widehat Z}{dE}\,\,\Big|\zeta_k \Big\rangle ,
\end{align}
where $|\zeta_k \rangle$ is the eigenstate given in Eq.(\ref{eq:Z(E)-eigenvalue-eq}).
The functions $\{F'_k(E)\}$ take the values
\begin{align}
\label{eq:values-of-derivative-Fk(E)-0}
  F'_k(E) &= 0 \hspace{1cm}\mbox{for} \,\,\,  E=E_i, \\
\label{eq:values-of-derivative-Fk(E)-2}
  F'_k(E) &= 2 \hspace{1cm}\mbox{for} \,\,\,  E=\varepsilon_j.
\end{align}
From the above properties of $\{F'_k(E)\}$  we see that the eigenvalues $\{E_i\}$ 
of $H$ can be obtained by calculating the solutions satisfying 
Eqs.(\ref{eq:Fk(E)=E}) and (\ref{eq:values-of-derivative-Fk(E)-0}) simultaneously.
A simple expression of the equation to be solved may be written as
\begin{align}
\label{eq:define-gk(E)}
  g_k(E) &= \left\{\frac{F_k(E) - E}{F_0} \right\}^2 + \{F'_k(E) \}^2\nonumber\\
              &= 0,
\end{align}
where $F_0$ is a parameter chosen suitably such that the two terms on the right-hand side
take values of the same order of  magnitude.

The solutions to Eq.(\ref{eq:define-gk(E)}) can be obtained by a graphical method. 
We define a function $f_k(E)$ as 
\begin{align}
\label{eq:define-fk(E)}
  f_k(E) = 
    \frac{1}{\{g_k(E)\}^2 +\Delta^2},               
\end{align}
where $\Delta$ is a small number.
The function $f_k(E)$ has the properties;
\begin{align}
\label{eq:limit-fk(E)-1}
  \lim_{E\rightarrow E_i}f_k(E) = \frac{1}{\Delta^2}             
\end{align}
and 
\begin{align}
\label{eq:limit-fk(E)-2}
  \lim_{E\rightarrow \varepsilon_j}f_k(E) = \frac{1}{4+\Delta^2},           
\end{align}
for the eigenvalues $E_i$ of $H$ and $\varepsilon_j$ of $QHQ$, 
which may be obvious from Eqs.(\ref{eq:Fk(E)=E})--(\ref{eq:define-gk(E)}). 
If the parameter $\Delta$ is taken to be small enough, the function $f_k(E)$ behaves 
like a resonance at $E=E_i$. 
By drawing the graph of $\{f_k(E), k=1,2,\cdots, d\}$ and finding resonance
positions, we obtain eigenvalues of $H$.
%
%
\subsection{Numerical calculation} 
In order to obtain some assessments of the present approach we study a model 
problem.
We start with a model Hamiltonian $H$ of which matrix elements are given by 
\begin{align}
\label{eq:define-random-matrix(E)}
  \langle i | H | j\rangle = (\alpha i +\beta i^2)\delta_{ij}  + \gamma x_{ij}
\end{align}
with  
\begin{align}
  x_{ij} &= 2\left\{ \sqrt{\sqrt{2} (i + j)} 
                               - \left [ \sqrt{\sqrt{2} (i + j)} \right ] \right \} - 1,
\end{align}
where $[X]$ is Gauss's symbol which means the integer part of a real number $X$. 
A set of $\{ x_{ij}\}$ are recognized as pseudo random numbers 
satisfying
\begin{align}
\label{eq:xij-range}
  -1 \leq x_{ij} \leq  1.             
\end{align}
The $\alpha ,\beta$ and $\gamma$ are the dimensionless parameters chosen suitably.
The total dimension of $H$ is taken to be $N_h = 100$.
As for the $P$ space we choose a two-dimensional space $(d=2)$ spanned by the two 
states which have the lowest and second lowest diagonal energies of $H$.
We here do not consider a case that  some of the eigenvalues $\{ y_k^{(m)}\}$ 
in Eqs.(\ref{eq:Yq1-operator-eigenvalue-eq}) and (\ref{eq:YQ(m-1)-operator-eigenvalue-eq})
become zero, because $\{ x_{ij}\}$ are pseudo random numbers and 
$H$ does not have any definite symmetry.
Therefore, the subspaces $\{ Q_k,\,\,\, k=1,2,\cdots, N_q \}$ are all $d$-dimensional
and the number of the subspace $\{ Q_k \}$ is given by 
$N_q = (N_h - 2)/2 = 49$.

We first calculate the $\widehat Q$ box and its energy derivatives $\widehat Q_1(E)$ 
and $\widehat Q_2(E)$ according to the continued-fraction method and the renormalized 
vertex method formulated in Sections III C and III D, respectively. 
We have confirmed numerically that the calculations on these two methods agree to each other.
With $\widehat Q(E)$, $\widehat Q_1(E)$ and $\widehat Q_2(E)$ we calculate 
the $\widehat Z$ box and its energy derivative $d\widehat Z (E)/dE$ according 
to Eqs.(\ref{eq:Z-box-definition}) and (\ref{eq:Z-box-differential}).

We next calculate the functions $F_k(E)$ and $F'_k(E)$ given in 
Eqs.(\ref{eq:Z(E)-eigenvalue-eq}) and (\ref{eq:define-derivative-Fk(E)}), respectively.
We finally obtain the functions $\{ f_k(E),\,\,\, k=1,2,\cdots, d \}$ and 
draw graphs of these functions.
Since the dimension of the $P$ space is taken to be $d=2$, we have two graphs 
of $f_1(E)$ and $f_2(E)$.
These graphs  are shown in Fig.{\ref{fig : resonance-position_fk(E)}.
From these figures we can specify the eigenvalues of $H$ as 
the resonance positions.
From Fig.{\ref{fig : resonance-position_fk(E)}, we can estimate four eigenvalues of $H$ 
on the interval $[0, 10]$.
%
\begin{figure}[!htb]
   \begin{center} 
   \includegraphics[width=10cm,keepaspectratio]{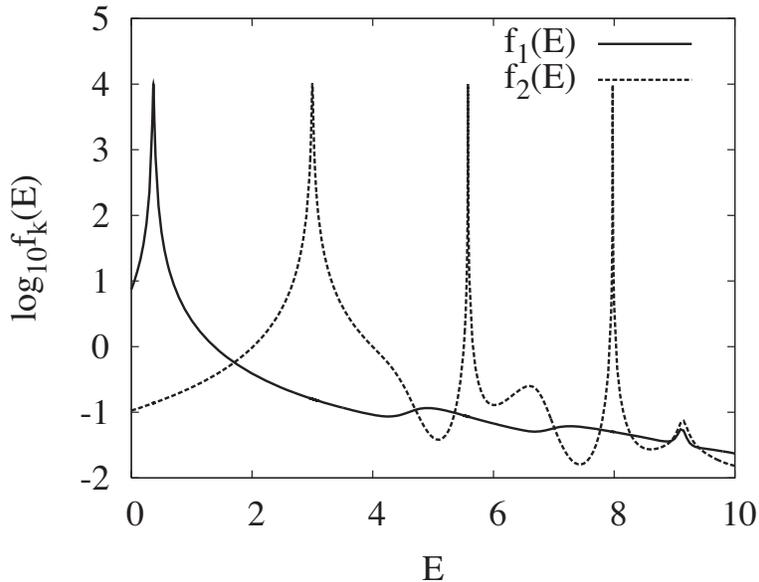} 
   \caption{Resonance-like behavior of the functions $f_1(E)$ and $f_2(E)$
in the case of $\alpha =1.2,\, \beta =0.2,\, \gamma =1.4,\,\Delta =10^{-2} $ 
and  $F_0=1.0$. 
The values of $E$ at the resonance positions correspond to the eigenvalues of the Hamiltonian. 
}
  \label{fig : resonance-position_fk(E)} 
  \end{center}
\end{figure}  
%

The accurate solution, namely $E_i$, can be obtained in the following way: 
We suppose that the solution $E_i$ lies on the interval $[a, b]$ and  there
are no other solutions on this interval.
The parabolic-interpolation method \cite{Bre73} is applied here.
If the difference $| E - E_i |$ is sufficiently small, the approximate form of 
$g_k(E)$ in Eq.(\ref{eq:define-gk(E)}) becomes  a parabolic function written as 
\begin{align}
\label{eq:parabolic-gk(E)}
  g_k(E) = \frac{1+\{   F_0 F''_k(E_i)\}^2}{F_0^2}( E - E_i )^2.        
\end{align}
Therefore we approximate $g_k(E)$ to be a parabolic function and solve the energy 
$E_i$ to give the minimum of $g_k(E)$. 
We note here that the parabolic function $A(x - \alpha )^2$ passing through two 
points $(a, g_k(a))$ and $(b, g_k(b))$ takes the minimum at the point $\alpha$ given by
\begin{align}
\label{eq:minimum-point-alpha}
  \alpha = \frac{a\sqrt{g_k(b)} + b\sqrt{g_k(a)}}{\sqrt{g_k(a)} + \sqrt{g_k(b)}} , 
\end{align}
where we have assumed $a<\alpha<b$. 
We utilize this fact to solve Eq.(\ref{eq:define-gk(E)}).

The calculation procedure employed in this numerical calculation is as follows: 
\begin{enumerate}
\item[(i)]
Determine an interval $[a, b]$ on which only one solution $E_i$ exists.\\
\item[(ii)]
Divide $[a,b]$ into equal intervals and define five points 
$(E_1, E_2, E_3, E_4, E_5)$ as
\begin{align}
\label{eq:five-points-Ek}
  E_k = a + (k-1) \Delta E, \hspace{1cm} 1\leq k \leq 5      
\end{align}
with $\Delta E = ( b - a )/4$.\\
\item[(iii)]
Consider all the intervals $[E_i, E_j]$ by selecting $E_i$ and $E_j$ among 
$\{E_1, E_2, \cdots, E_5\}$ and calculate
\begin{align}
\label{eq:energy-Eij}
  E_{ij} = \frac{E_i \sqrt{g_k(E_j)} + E_j \sqrt{g_k(E_i)}}
                        {\sqrt{g_k(E_j)} + \sqrt{g_k(E_i)}}.     
\end{align}
\item[(iv)]
There are ten combinations of the energies $\{E_{ij}\}$. 
Arrange $\{E_{ij}\}$ in order of energy and write them as 
$u_1 < u_2 < \cdots < u_{10}$.\\
\item[(v)]
Calculate the values $\{g_k(u_n),\,n=1,2,\cdots,10\}$ and find the minimum 
$g_k(u_m)$ as shown in Fig.{\ref{fig : new-interval}}.
We determine a new interval $[a, b]$ according to 
\begin{align}
\label{eq:new-interval}
 a &= u_{m-1}, \,\,\,\, b =u_{m+1} \hspace{1cm} \mbox{if}\,\,\,\,2\leq m \leq 9,    \nonumber\\
 a &= a, \,\, \,\,b = u_2 \hspace{1cm} \mbox{if}\,\,\,\, m=1,    \nonumber\\
 a &= u_9,  \,\,\,\, b =b \hspace{1cm} \mbox{if}\,\,\,\, m=10.   
\end{align}
\item[(vi)]
Repeat the procedure until the convergence, $|g_k(u_m)| < \delta$, is attained
for an appropriate small number $\delta$.
\end{enumerate}
%
\begin{figure}[!htb]
   \begin{center} 
\includegraphics[width=10cm,keepaspectratio]{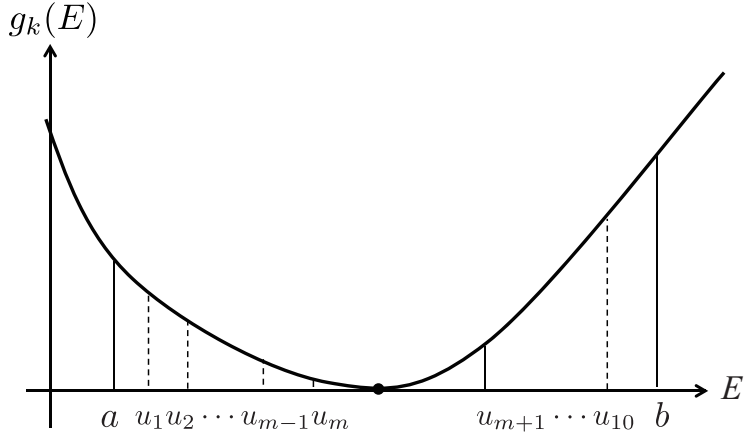} 
   \caption{
Illustration of determining a new interval for finding the minimum point of the function $g_k(E)$. 
If $g_k(u_m) $ is the minimum value among 
   $\{g_k(u_n),\,n=1,2,\cdots,10\}$, then the new interval is given by 
   $[a, b] = [u_{m-1}, u_{m+1}]$.}
  \label{fig : new-interval} 
  \end{center}
\end{figure}  

In Table \ref{table:lowest-two-eigenvalues} we show the results for the lowest two eigenvalues 
of $H$ calculated by the above mentioned parabolic-interpolation method.
The convergence is markedly fast.
With three times of the changes of the interval $[a, b]$, convergence is reached
with accuracy better than 10 decimal places.
%
\begin{table}[!htb]
  \caption{Correct digits of the lowest two eigenvalues of $H$ calculated by 
the parabolic-interpolation method. The parameters $\alpha, \beta$, and $\gamma$ are taken  
as the same as in Fig.\ref{fig : resonance-position_fk(E)}. Initial intervals are taken to be 
$[a, b]=[0.0,1.0]$ and $[2.5,3.5]$ for $E_1$ and $E_2$, respectively.
}
  \label{table:lowest-two-eigenvalues}
\begin{center}
\begin{tabular}{ccl} \hline \hline
  $E_i$           &   \,\,No. of repeats\,\,   &  Calculated value     \\ \hline
  $E_1$         &        1             &      0.365      \\
                      &         2          &      0.365550      \\
                      &         3          &      0.365550151994574      \\ \hline
 $E_2$          &        1              &      2.999       \\
                      &          2          &      2.9994240     \\
                      &          3          &      2.99942408730107      \\ \hline\hline
\end{tabular}
 \end{center}
 \end{table}
%

As has been shown in Eq.(\ref{eq:Q_box_solution_inverse_propagator_final}), 
the $\widehat Q$ box is given 
by a sum over the number $k$.  
In this model calculation the maximum number of $k$ is equal to $N_q = 49$.   
 Introducing a number $K_{\rm max}$, we consider a truncation as $k \leq K_{\rm max}$ 
in the calculation of the $\widehat Q$ box in Eq.(\ref{eq:Q_box_solution_inverse_propagator_final}). 
It would be interesting to examine the dependence of the calculated 
eigenvalues of $H$ on $K_{\rm max}$.  
 The results are shown in Figs.\ref{fig : E1-Kmax} and \ref{fig : E2-Kmax}.
 It is clear that, as $K_{\rm max}$ approaches to $N_q$=49, the eigenvalues converge 
 to the exact values.   
 These results suggest a possibility of introducing a new way of truncation in the series expansion 
 for the $\widehat Q$ box, instead of making it according to the magnitude of energies of intermediate 
states as in the usual perturbative calculations. 
%
\begin{figure}[!htb]
   \begin{center} 
\includegraphics[width=11cm,keepaspectratio]{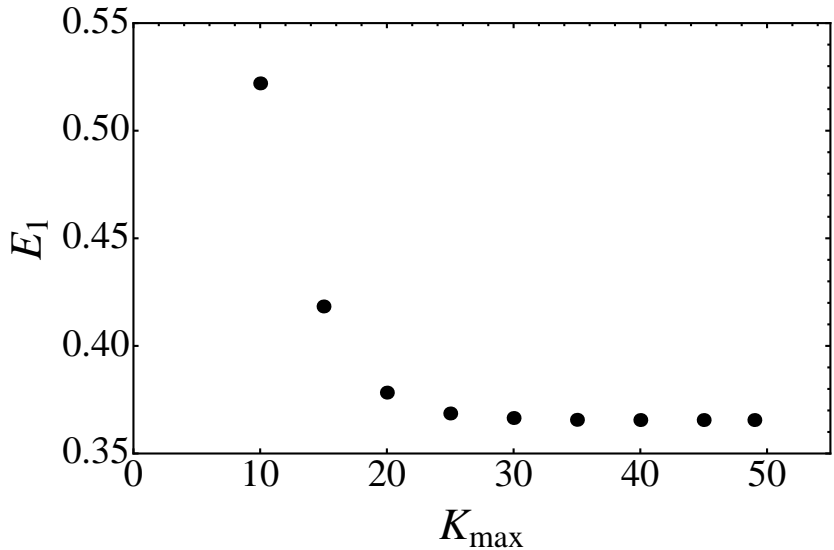} 
   \caption{Convergence of $E_1$ as a function of $K_{\rm max}$.
The $K_{\rm max}$ denotes the block dimension which means the number of the subspaces 
\{$Q_k$, $k$=1,2,$\cdots$,$K_{\rm max}$\} taken into calculation. In this model calculation 
$K_{\rm max}$ is in the range $1\le K_{\rm max} \le49$. 
The exact value of $E_1$ is 0.36555$\cdots$ as given in Table \ref{table:lowest-two-eigenvalues}.
}
  \label{fig : E1-Kmax} 
  \end{center}
\end{figure}  
%
\begin{figure}[!htb]
   \begin{center} 
     \includegraphics[width=11cm,keepaspectratio]{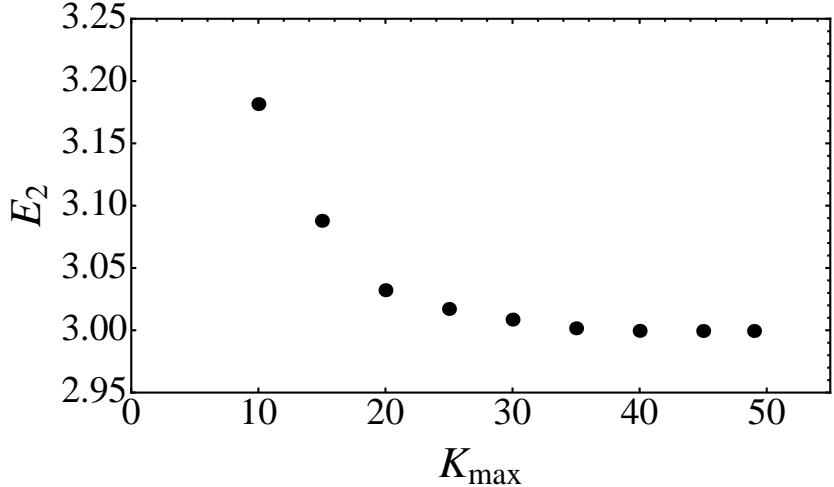}   
   \caption{Convergence of $E_2$ as a function of $K_{\rm max}$.
The exact value of $E_2$ is 2.9994$\cdots$. 
Other notations are the same as in Fig.\ref{fig : E1-Kmax}.
}
  \label{fig : E2-Kmax} 
  \end{center}
\end{figure} 
%
%
\section{Concluding Remarks}
We have proposed a new approach to the effective interaction and/or Hamiltonian 
acting within a model space $P$. 
In the present stage of the effective-interaction theory one of the central 
problems has been how to calculate accurately the $\widehat Q$ box which has been used 
as a building block of the formulation.  
The main concern of the present study has been to derive a new method of calculating 
the $\widehat Q$ box as accurately as possible even if the original Hamiltonian $H$ is given 
in a huge-dimensional space.

The formulation consists of two steps:  First one is to transform a given Hamiltonian $H$ 
to a block-tridiagonal form by dividing the complementary space $Q$ of the $P$ space 
into subspaces $\{Q_k,\,\, k=1,2,\cdots\}$ with tractable dimensions.
If the subspaces are chosen suitably the Hamiltonian is transformed to a block-tridiagonal 
form. 
With the Hamiltonian thus transformed, the next step is to derive coupled equations 
for determining the $\widehat Q$ box.  
By solving these coupled equations we have proved that the $\widehat Q$ box can be 
represented in two ways : 
The first one is that the $\widehat Q$ box is expanded into a form of  continued 
fraction in terms of the submatrices  which are the elements of the block-tridiagonalized 
Hamiltonian.  
It has been proved that if a quantum system can be well described by a Hamiltonian given 
in a finite dimensional space, the continued fraction can be reduced to only one
term with a renormalized propagator which can be calculated by using 
a descending recurrence relation.  
The other solution is obtained by using ascending recurrence relations for solving 
the coupled equations. 
The resultant $\widehat Q$ box can be shown to be given by only two terms such 
as $PHP$ and a sum of second-order terms with respect to renormalized vertices 
and propagators.  
This reduction of the $\widehat Q$ box has clarified that there exists a method of determining 
renormalized vertices and propagators such that the $\widehat Q$ box can be given 
by a sum of terms up to second order.
 
Given the $\widehat Q$ box, we have applied the $\widehat Z$-box method for solving 
the eigenvalue problem of a Hamiltonian $H$. 
We have introduced functions of energy variable $E$ as $\{f_k(E),\,\, k=1,2,\cdots,d\}$ 
such that $f_k(E)$ behaves like a resonance at $E=E_i$ which is one of the eigenvalues of $H$. 
Here the number $d$ is the dimension of the model space.  
In this approach the eigenvalues of $H$ can be given by the resonance positions of the functions 
$\{f_k(E),\,\,k=1,2,\cdots,d\}$. 
This approach enables us to solve the eigenvalue equation of $H$ in a graphical way. 
We  here emphasize that there would be an applicability of the present approach to solving 
the eigenvalue problem for a Hamiltonian given in a huge-dimensional shell-model space, 
because the calculation procedures include only manipulations of matrices with dimensions 
less-than-or-equal-to $d$. 

In order to assess the present method we have made a test calculation by
introducing a 100$\times$100 model Hamiltonian.  
We have performed the calculation of the $\widehat Q$ box by employing two methods, namely, 
the continued-fraction expansion and the expansion with the renormalized vertices and 
propagators.  We have confirmed that both the two methods have reproduced the exact 
eigenvalues of the original Hamiltonian $H$.

The present  non-perturbative method would have another possibility of application 
to the derivation of the effective interaction to be used in the shell-model calculations. 
The reduction of the $\widehat Q$ box to simple second-order diagrams may attain 
a simplification of the calculation of the effective interaction.  
We here note, however, that the present study is based essentially on the algebraic approach 
to the effective Hamiltonian. 
For the calculation of the effective interaction among valence particles outside the core,  
it is necessary to represent the $\widehat Q$ box in terms of linked diagrams.
 A general relation is not clear between the present approach and the linked-and-folded-diagram theory.
Therefore, this formal relation is an important problem to be clarified.
%
\begin{acknowledgments}

The authors are grateful to T.T.S.~Kuo for his continuous interest in this work and encouragement. 
We thank T.~Mizusaki and K.~Takayanagi for their instructive  discussions.

\end{acknowledgments}
%
\appendix
\section{Derivatives of the $\widehat Q$ box in Eq.(\ref{eq:Q-box-tilde-e-(1)}) }
The first and second derivatives of the $\widehat Q$ box are given, respectively, by  
\begin{align}
\label{eq:derivative-Q}
 \frac{d\widehat Q(E)}{dE} = 
     &-PHQ_1 \frac{1}{\widetilde e_1(E) } \widetilde k_1(E) \frac{1}{\widetilde e_1(E) } Q_1HP, \\
\label{eq:2nd-derivative-Q}
 \frac{d^2\widehat Q(E)}{dE^2} = 
     & \,\,\,2 PHQ_1 \frac{1}{\widetilde e_1(E) } \widetilde k_1(E) \frac{1}{\widetilde e_1(E) } 
           \widetilde k_1(E) \frac{1}{\widetilde e_1(E) }Q_1HP\cr
     &-  PHQ_1\frac{1}{\widetilde e_1(E) }\widetilde l_1(E) \frac{1}{\widetilde e_1(E) } Q_1HP.
\end{align}
Here $\widetilde e_1(E), \widetilde k_1(E)$ and $\widetilde l_1(E)$ are given 
through
 the following recurrence relations:
We consider the energy derivative of $\widetilde e_n(E)$ in Eq.(\ref{eq:tilde-e-(n-1)}) and write as
\begin{align}
\label{eq:define-tilde-k-(n)}
\widetilde k_n(E) = \frac{d\widetilde e_n(E)}{dE}.
\end{align}
Noting a relation 
\begin{align}
\label{eq:derivative-inverse-tilde-e-(n)}
 \frac{d}{dE}\Big\{\frac{1}{\widetilde e_n(E)}\big\}
   &= -\frac{1}{\widetilde e_n(E)} \frac{d\widetilde e_n(E)}{dE} \frac{1}{\widetilde e_n(E)}\cr
   &= -\frac{1}{\widetilde e_n(E)} \widetilde k_n(E) \frac{1}{\widetilde e_n(E)},
\end{align}
we can derive 
\begin{align}
\label{eq:tilde-k-(n)}
 \widetilde k_n(E) = 
     Q_n + H_{n,n+1} \frac{1}{\widetilde e_{n+1}(E) } \widetilde k_{n+1}(E)  
     \frac{1}{\widetilde e_{n+1}(E)} H_{n+1,n} ,
\end{align}
where we have used the energy derivative of $e_n(E)$ in Eq.(\ref{eq:e-1(E)})
\begin{align}
\label{eq:derivative-e-(n)}
 \frac{d e_n(E)}{dE} = Q_m.
\end{align}
The $H_{n,n+1}$ and $H_{n+1,n}$ are defined in Eq.(\ref{eq:Hij}).
For the maximum number of $n$, denoted by $N$, $\widetilde k_N(E)$ is given by
\begin{align}
\label{eq:tilde-k-(N)}
 \widetilde k_N(E) &= \frac{d \widetilde e_N(E)}{dE}\cr
                        &=  Q_N,
   \end{align}
which is derived from Eq.(\ref{eq:tilde-e-(N)}) for $\widetilde e_N(E)$.    
Starting with $\widetilde k_N(E)$, the recurrence relation determines a sequence
$\widetilde k_N(E)$, $\widetilde k_{N-1}(E)$, $\cdots$, $\widetilde k_1(E)$.

We write the second derivative of $\widetilde e_n(E)$ as 
\begin{align}
\label{eq:tilde-l-(n)}
 \widetilde l_n(E) &= \frac{d^2 \widetilde e_n(E)}{dE^2}\cr
                      &= \frac{d \widetilde k_n(E)}{dE}.
\end{align}
From Eq.(\ref{eq:tilde-k-(n)}) for $\{\widetilde k_n(E)\}$ a recurrence formula for 
$\{\widetilde l_n(E)\}$ can be derived as
\begin{align}
\label{eq:recurrence-tilde-l-(n)}
\widetilde l_n(E) = 
     &- 2 H_{n,n+1} \frac{1}{\widetilde e_{n+1}(E)} \widetilde k_{n+1}(E) 
           \frac{1}{\widetilde e_{n+1}(E) } \widetilde k_{n+1}(E) \frac{1}{\widetilde e_{n+1}(E) }
          H_{n+1,n}\cr
     &+ H_{n,n+1} \frac{1}{\widetilde e_{n+1}(E)} \widetilde l_{n+1}(E) 
           \frac{1}{\widetilde e_{n+1}(E) } H_{n+1,n}.
\end{align}
For the maximum number $n=N$ the $\widetilde l_N(E) $ is given, from 
Eqs.(\ref{eq:tilde-k-(N)}) and (\ref{eq:tilde-l-(n)}), by
\begin{align}
\label{eq:tilde-l-(N)=0}
 \widetilde l_N(E)=0.
 \end{align}
The recurrence formula Eq.(\ref{eq:recurrence-tilde-l-(n)}) determines a sequence 
$\widetilde l_N(E), \widetilde l_{N-1}(E), \cdots , \widetilde l_1(E)$.
Substituting the operators $\widetilde e_1(E), \widetilde k_1(E)$ and 
$\widetilde l_1(E)$ into Eqs.(\ref{eq:derivative-Q}) and (\ref{eq:2nd-derivative-Q}) 
the first and second derivatives of the $\widehat Q$ box can be calculated. 

Here it should be noted that the first and second derivatives of the 
$\widehat Q$ box can be expressed by using only small-dimensional matrices. 
The $\widetilde e_1(E), \widetilde k_1(E)$ and $\widetilde l_1(E)$ are the operators on the 
subspace $Q_1$ which are represented by $d_1$$\times$$d_1$ matrices.
The operator $PHQ_1$ is a mapping between the $P$ and $Q_1$ spaces 
and has a $d$$\times$$d_1$ matrix representation.
%
%
\section{Derivatives of the $\widehat Q$ box 
in Eq.(\ref{eq:Q_box_solution_inverse_propagator_final})}
We derive the first and second 
derivatives of the $\widehat Q$ box with respect to energy variable $E$ as 
\begin{align}
\label{eq:dQ/dE}
\frac{d\widehat Q(E)}{dE}
&=\sum_{k=1}^\infty\{(\overline H'_k(E)\lambda_k(E)\overline H_k^\dagger
(E)
+ {\rm h.c.})
+ \overline H_k(E)\lambda '_k(E)\overline H_k^\dagger (E) \}, \\
\label{eq:d2Q/dE2}
\frac{d^2\widehat Q(E)}{dE^2}
&=\sum_{k=1}^\infty\{2\overline H'_k(E)\lambda_k(E)\overline H_k'^{\dagger} (E)
    + (\overline H''_k(E)\lambda_k(E)\overline H_k^\dagger (E) + {\rm h.c.})
       \notag \\
&\hspace{5mm}+2(\overline H'_k(E)\lambda '_k(E)\overline H_k^\dagger (E)
    + {\rm h.c.})
    + \overline H_k(E)\lambda ''_k(E)\overline H_k^\dagger (E) \}
\end{align}
with 
\begin{align}
\label{eq:overlineH'k(E)}
\overline H'_k(E) &= \frac{d\widehat H_k(E)}{dE},\\
\label{eq:overlineH''k(E)}
\overline H''_k(E) &= \frac{d^2\widehat H_k(E)}{dE^2},\\
\label{eq:lamk(E)}
\lambda_k(E) &= \frac{1}{\overline e_k(E)},\\
\label{eq:lam'k(E)}
\lambda '_k(E) &= \frac{d\lambda_k(E)}{dE}\cr
      &= -\frac{1}{\overline e_k(E)}\frac{d\overline e_k(E)}{dE}\frac{1}{\overline e_k(E)},\\
\label{eq:lam''k(E)}      
\lambda ''_k(E) 
&= \frac{d^2\lambda_k(E)}{dE^2}\cr
&= 2\frac{1}{\overline e_k(E)}\frac{d\overline e_k(E)}{dE}
   \frac{1}{\overline e_k(E)}\frac{d\overline e_k(E)}{dE}\frac{1}{\overline e_k(E)}
   -\frac{1}{\overline e_k(E)}\frac{d^2\overline e_k(E)}{dE^2}\frac{1}{\overline  
   e_k(E)}.
\end{align}
These expressions indicate that the calculation of the derivatives of $\widehat Q(E)$ is reduced 
to that of $\{\overline H_k(E)\}$, $\{\lambda_k(E)\}$ and their derivatives; 
$\{\overline H_k(E)\}$ are given through the recurrence relation Eq.(\ref{eq:overline_H_k+1}) 
and accordingly its derivatives are 
\begin{align}
\label{eq:overlineH'n+1(E)}
\overline H'_k(E)
	        &= \{\overline H'_{k-1}(E) \lambda_{k-1}(E)  
                                      +  \overline H_{k-1}(E) \lambda '_{k-1}(E)  \}H_{k-1,k},\\
\label{eq:overlineH''n+1(E)}
\overline H''_{k}(E) 
        &= \{\overline H''_{k-1}(E) \lambda_{k-1}(E) 
             + 2\overline H'_{k-1}(E) \lambda '_{k-1}(E)
             +  \overline H_{k-1}(E) \lambda ''_{k-1}(E) \}H_{k-1,k},
\end{align}
and $\{\lambda_k(E)\}$ is given by
\begin{align}
\label{eq:lamn+1(E)}
\lambda_{k}(E) &= \{e_{k}(E) - H_{k,k-1}\lambda_{k-1}(E)H_{k-1,k}\}^{-1}
\end{align}
from Eqs.(\ref{eq:inverse_propagator}) and (\ref{eq:lamk(E)}), and accordingly 
its derivatives are 
\begin{align}
\label{eq:lam'n+1(E)}
\lambda '_{k}(E) = &
     -\lambda^{2}_{k}(E)+\lambda_{k}(E) H_{k,k-1}\lambda '_{k-1}(E) H_{k-1,k}\lambda_{k}(E),
     \\ \label{eq:lam''n+1(E)} 
\lambda ''_{k}(E) = &
     -(\lambda '_{k}(E)\lambda_{k}(E)+{\rm h.c.})+\{ \lambda '_{k}(E) H_{k,k-1}\lambda '_{k-1}(E)
H_{k-1,k}\lambda_{k}(E) + {\rm h.c.} \} \notag \\ &+ \lambda_{k}(E) H_{k,k-1} \lambda ''_{k-1}(E)
H_{k-1,k}\lambda_{k}(E).
\end{align}
Their initial values are 
given in
Eq.(\ref{eq:overline_H_1}) for $\{\overline H_k(E)\}$, 
$\overline H'_1(E)=\overline H''_1(E)=0$ and 
\begin{align}
\label{eq:lam1}
\lambda_{1}(E) &= \frac{1}{e_1(E)}\cr
                            &= (E - Q_1HQ_1)^{-1},\\
 \label{eq:lam'1}
\lambda '_{1}(E) &= -(E - Q_1HQ_1)^{-2},\\
\label{eq:lam''1}
\lambda ''_{1}(E) &= 2(E - Q_1HQ_1)^{-3} .
\end{align}

The way of calculating the derivatives of the $\widehat Q$ box is summarized 
as follows: First, $\{\lambda_k(E)\}$ is calculated by Eq.(\ref{eq:lamn+1(E)}), then 
their derivatives by Eqs.(\ref{eq:lam'n+1(E)}) and (\ref{eq:lam''n+1(E)}). 
Next $\{\overline H'_k(E)\}$ and $\{\overline H''_k(E)\}$ by 
Eqs.(\ref{eq:overlineH'n+1(E)}) and (\ref{eq:overlineH''n+1(E)}), finally we obtain 
the derivatives of the $\widehat Q$ box. 
Here $\lambda_k(E)$, $\lambda '_k(E)$, and $\lambda ''_k(E)$ are $d_k$$\times$$ d_k$ 
matrices, while $\overline H_k(E)$, $\overline H'_k(E)$, and 
$\overline H''_k(E)$ are $d$$\times$$ d_k$ matrices. 

%


\begin{thebibliography}{99}
\bibitem{PRWC01} S.~C.~Pieper, V.~R.~Pandharipande, R.~B.~Wiringa, and  J.~Carlson, 
Phys.~Rev.~C~{\bf 64}, 014001 (2001).
\bibitem{PW01} S.~C.~Pieper and R.~B.~Wiringa, Annu.~Rev.~Nucl.~Part.~Sci.~{\bf 51}, 53 (2001).
\bibitem{NVB00-2} P.~Navr\'atil, J.~P.~Vary, B.~R.~Barrett, Phys.~Rev.~C~{\bf62}, 
054311 (2000).
\bibitem{LBKNSV08} A.~F.~Lisetskiy, B.~R.~Barrett, M.~K.~G.~Kruse, P.~Navratil, I.~Stetcu, J.~P.~Vary, 
Phys.~Rev.~C~{\bf78}, 044302 (2008).
\bibitem{LO12} W.~Leidemann, G.~Orlandini, arXiv:nucl-th/1204.4617v1.
\bibitem{BM07} R.~J.~Bartlett, M.~Musial, Rev.~Mod.~Phys.~{\bf79}, 291 (2007).
\bibitem{HPDH08} G.~Hagen, T.~Papenbrock, D.~J.~Dean, M.~Hjorth-Jensen, Phys.~Rev.~Lett.~{\bf101}, 
092502 (2008).
\bibitem{HPDH09} G.~Hagen, T.~Papenbrock, D.~J.~Dean, M.~Hjorth-Jensen, B.~Velamur Asokan, 
Phys.~Rev.~C~{\bf80}, 021306(R) (2009).
\bibitem{SOK87} K.~Suzuki, R.~Okamoto, H.~Kumagai, Phys.~Rev.~C~{\bf 36}, 804 (1987).
\bibitem{FOS04} S.~Fujii, R.~Okamoto, K.~Suzuki, Phys.~Rev.~C~{\bf 69}, 034328 (2004).
\bibitem{FOS09} S.~Fujii, R.~Okamoto, K.~Suzuki, Phys.~Rev.~Lett.~{\bf 103}, 182501 (2009).
\bibitem{Lan50} C.~Lanczos, J.~Res.~Nat.~Bur.~Stand.~{\bf 45}, 255 (1950).
\bibitem{GL96} G.~H.~Golub, C.~F.~Van Loan, {\it Matrix Computations} 
(The Johns Hopkins University Press, 1996).
\bibitem{MKHS10} T.~Mizusaki, K.~Kaneko, M.~Honma, T.~Sakurai, Phys.~Rev.~C~{\bf 82},
024310 (2010).
\bibitem{Tal63} I.~Talmi,  Rev. ~Mod.~ Phys.~{\bf 34}, 704 (1963).
\bibitem{MBZ64} J.~D.~ McCullen, B.~F.~Bayman, L.~Zamick, Phys. Rev.~{\bf 134}, B515 (1964).
\bibitem{Tal03} I.~Talmi, Adv.~Nucl.~Phys.~{\bf 27}, 1 (2003).
\bibitem{DTW62} J.~F.~Dawson, I.~Talmi, J.~D.~Walecka, Ann.~Phys.~{\bf 18}, 339 (1962).
\bibitem{Ber65} G.~F.~Bertsch, Nucl.~Phys.~{\bf 74}, 234 (1965).
\bibitem{KB66} T.~T.~S.~Kuo, G.~E.~Brown, Nucl.~Phys.~{\bf 85}, 40 (1966).
\bibitem{BK70} B.~R.~Barrett, M.~W.~Kirson, Nucl.~Phys.~{\bf A148}, 145 (1970).
\bibitem{HSK76} H.~M.~Hofmann, Y.~Starkand, M.~W.~Kirson,  Nucl.~Phys.~{\bf A266}, 138 (1976).
\bibitem{AM79} N.~Ayoub, H.~A.~Mavromatis, Nucl.~Phys.~{\bf A323}, 125 (1979).
\bibitem{Kir74} M.~W.~Kirson, Ann.~Phys.~{\bf 82}, 345 (1974).
\bibitem{Kir71} M.~W.~Kirson, Ann.~Phys.~{\bf 68}, 556 (1971).
\bibitem{BB73} S.~Babu, G.~E.~Brown, Ann.~Phys.~{\bf 78}, 1 (1973).
\bibitem{HHKBB05} J.~D.~Holt, J.~W.~Holt, T.~T.~S.~Kuo, G.~E.~Brown, S.~K.~Bogner,
                 Phys.~Rev.~{\bf C72}, 041304(R) (2005).
\bibitem{KLR71} T.~T.~S.~Kuo, S.~Y.~Lee, K.~F.~Ratcliff, Nucl.~Phys.~A~{\bf 176}, 65 (1971).
\bibitem{Bra67} B.~H.~Brandow, Rev.~Mod.~Phys.~{\bf 39}, 771 (1967).
\bibitem{BK73} B.~R.~Barrett, M.~W.~Kirson, Adv.~Nucl.~Phys,~{\bf 6}, 219 (1973).
\bibitem{EO77} P.~J.~Ellis and E.~Osnes, Rev.~Mod.~Phys.~{\bf 49}, 777 (1977).
\bibitem{KO90} T.~T.~S.~Kuo, E.~Osnes,  {\it Lecture Notes in Physics}, Vol.~364
(Springer-Verlag, New York, 1990).
\bibitem{And87} F.~Andreozzi, Phys.~Rev.~C~{\bf 54}, 684 (1996).
\bibitem{RW12} D.~J.~Rowe, J.~L.~Wood, {\it FUNDAMENTALS of NUCLEAR MODELS 
-- Foundational Models} (World Scientific, 2012).
\bibitem{CCGIK09} L.~Coraggio, A.~Covello, A.~Gargano, N.~Itaco, T.~T.~S.~Kuo, Prog.~Part.~Nucl.~Phys.~{\bf 62}, 135 (2009).
\bibitem{CCGIK12} L.~Coraggio, A.~Covello, A.~Gargano, N.~Itaco, T.~T.~S.~Kuo, Ann.~Phys.~{\bf 327}, 2125 (2012).
\bibitem{VNUS01} K.~Varga, P.~Navr\'atil, J.~Usukura, Y.~Suzuki, Phys.~Rev.~B~{\bf 63}, 
205308 (2001).
\bibitem{PHHKP11} M.~Pedersen~Lohne, G.~Hagen, M.~Hjorth-Jensen, S.~Kvaal, F.~Pederiva, 
Phys.~Rev.~B~{\bm 84}, 115302 (2011).
\bibitem{CFAR09} J.~Christensson, C.~Forss\'en, S.~\AA berg, S.~M.~Reimann,
Phys.~Rev.~A~{\bf 79}, 012707 (2009).
\bibitem{HKO95} M.~Hjorth-Jensen, T.~T.~S. Kuo, E.~Osnes, Phys.~Rep.~{\bf 261}, 
125 (1995).
\bibitem{DEHKO04} D.~J.~Dean, T.~Engeland, M.~Hjorth-Jensen, M.~Kartamychev, E.~Osnes, 
Prog.~Part.~Nucl.~Phys.~{\bf 53}, 419 (2004).
\bibitem{BH58} C.~Bloch, J.~Horowitz, Nucl.~Phys.~{\bf 8}, 91 (1958).
\bibitem{Fesh62} H.~Feshbach, Ann. Phys. N. Y. {\bf 19}, 287(1962).
\bibitem{LM86} I.~Lindgren, J.~Morrison, {\it Atomic Many-Body Theory}, second edition
(Springer-Verlag, 1986).
\bibitem{WH12} S.~Wilson, I.~Huba$\check{\rm c}$, {\it Brillouin-Wigner Methods for Many-Body Systems}, 
Progress in Theoretical Chemistry and Physics (Springer-Verlag, 2012).
\bibitem{KK74} E.~M.~Krenciglowa, T.~T.~S.~Kuo,  Nucl.~Phys.~A~{\bf 235}, 171 (1974).
\bibitem{LS80} S.~Y.~Lee, K.~Suzuki, Phys.~Lett. {\bf B91} 173 (1980).
\bibitem{SL80} K.~Suzuki, S.~Y.~Lee, Prog.~Theor.~Phys.~{\bf 64}, 2091 (1980).
\bibitem{SOEK94} K.~Suzuki, R.~Okamoto, P.~J.~Ellis, T.~T.~S.~Kuo,
Nucl.~Phys.~A~{\bf 567}, 570 (1994).
\bibitem{Tak11b} K.~Takayanagi, Nucl.~Phys.~A~{\bf 864}, 91 (2011).
\bibitem{SOKF11} K.~Suzuki, R.~Okamoto, H.~Kumagai, S.~Fujii, Phys.~Rev.~C~{\bf 83}, 
024304 (2011).
\bibitem{Oku54} S.~Okubo, Prog.~Theor.~Phys.~{\bf 12}, 603 (1954).
\bibitem{JT80} W.~B.~Jones, W.~J.~Thron, {\it Continued Fractions}, 
Encyclopedia of Mathematics and its Applications, Vol.~11 (Addison-Wesley, 1980).
\bibitem{Bre73} R.~P.~Brent, {\it ALGORITHMS FOR MINIMIZATION WITHOUT DERIVATIVES} 
(Dover, 1973).
\bibitem{SKS83} J.~Shurpin, T.~T.~S.~Kuo, D.~Strottman, Nucl.~Phys.~{\bf A408}, 310 (1983).
\bibitem{ACCGKLP12} F.~Andreozzi, L.~Coraggio, A.~Covello, A.~Gargano, T.~T.~S.~Kuo,
   Z.~B.~Li, A.~Porrino, Phys.~Rev.{\bf C54}, 1636 (1996).


\end{thebibliography}
\end{document}